\documentclass[10pt,twocolumn,showpacs,nofootinbib,aps,prd]{revtex4-2}
\usepackage{natbib}
\usepackage{mathrsfs}
\usepackage{latexsym,amsfonts,amsmath,amssymb}
\usepackage{hyperref}
\usepackage{bbm}
\usepackage{empheq}
\usepackage{graphicx}
\usepackage{color}
\usepackage{caption}
\usepackage{subcaption}
\usepackage[normalem]{ulem}
\usepackage{comment}
\usepackage{url}
\usepackage{slashed}
\usepackage{extarrows}
\usepackage{amsmath}
\usepackage{mathtools}
\usepackage{float}
\usepackage{cancel}
\usepackage{amsxtra,graphics,epsfig,bm,tikz,xfrac}
\usepackage{array}

\newcommand\NN{\mathcal{N}}
\newcommand\AAA{\mathcal{A}}
\newcommand\BB{\mathcal{B}}

\newcolumntype{P}[1]{>{\centering\arraybackslash}p{#1}}
\begin{document}
\title{Variant dilaton Weyl Multiplet for $\mathcal{N}=3$ conformal supergravity
in four dimensions}
\author{Soumya Adhikari} 
\email{ssoumya.a012@gmail.com}
\affiliation{
Indian Institute of Science Education and Research Thiruvananthapuram}

\author{Aravind Aikot} 
\email{arvd1719@alumni.iisertvm.ac.in}
\affiliation{
Indian Institute of Science Education and Research Thiruvananthapuram}

\author{Madhu Mishra} 
\email{madhu.mishra@apctp.org}
\affiliation{
Asia Pacific Center for Theoretical Physics, Postech, Pohang 37673, Korea }

\author{Bindusar Sahoo} 
\email{bsahoo@iisertvm.ac.in}
\affiliation{
Indian Institute of Science Education and Research Thiruvananthapuram}
\begin{abstract} 
We construct a new dilaton Weyl multiplet for $\mathcal{N}=3$ conformal supergravity in four dimensions. The R-symmetry realized on this dilaton Weyl multiplet is $SU(2) \times U(1) \times U(1)$. The construction follows a two-step procedure. Firstly, two on-shell vector multiplets are coupled to the standard Weyl multiplet. Secondly, using the field equations of the vector multiplets, some of the auxiliary fields of the standard Weyl multiplet are solved in terms of the fields belonging to the vector multiplets and some dual gauge fields. The remaining fields of the standard Weyl multiplet combine with the vector multiplet fields and the dual gauge fields to constitute the new dilaton Weyl multiplet.
\end{abstract}    
\allowdisplaybreaks

\maketitle

\section{Introduction} 
The study of supergravity is motivated by its potential to reconcile general relativity with the principles of supersymmetry, thereby unifying the fundamental forces of nature and offering insights into the behavior of gravity at the quantum level. The local symmetries realized in a supergravity theory are diffeomorphism, which is a space-time symmetry and other local symmetries such as local Lorentz transformation ($M_{ab}$) and local supersymmetry ($Q_a^i$) which are realized on the tangent space. The corresponding gauge fields are $e_\mu ^a$, $\omega_{\mu}{}^{ab}$, and $\psi_{\mu}^i$, respectively. The gauge fields $e_{\mu}^{a}$ and $\omega_{\mu}^{ab}$ have the standard interpretation of vielbein and spin connection, respectively and are related to each other via a supersymmetric extension of the Cartan's first structure equation. Since the local symmetries realized on such a supergravity theory is based on a super-Poincar{\'e} algebra, we often refer to such a supergravity theory as Poincar{\'e} supergravity.

For various purpose, one needs to construct matter coupled Poincar{\'e} supergravity theories including higher derivative corrections. Such a construction is greatly facilitated using the superconformal tensor calculus approach. The superconformal tensor calculus approach relies on the gauge-equivalence principle, which states that Poincar{\'e} supergravity arises under a specific gauge choice of conformal supergravity. As the name suggests, the local symmetries of conformal supergravity is based on the superconformal algebra, which contains the super-Poincar{\'e} algebra as a subalgebra together with some extra symmetry generators such as dilatation, special conformal transformation, R-symmetries and a special or S-supersymmetry. By gauge-fixing the extra symmetries using certain matter multiplets as compensators, conformal supergravity reduces to Poincar\'e supergravity \cite{Ferrara:1977ij, Freedman:2012zz}. This approach takes advantage of the larger set of symmetries in conformal supergravity theories, allowing for the organization of fields into shorter multiplets thereby making the construction of conformal supergravity systematic and tractable.

The most important ingredient of a conformal supergravity theory is the Weyl multiplet, which contains all the gauge fields associated with the superconformal algebra. In an extended conformal supergravity theory, the Weyl multiplet also contains several covariant matter fields to ensure the off-shell closure of the superconformal algebra. For a given dimension and number of supersymmetries, there can exist two kinds of Weyl multiplets: the standard Weyl multiplet and the dilaton Weyl multiplet\footnote{A notable exception arises in five-dimensional $\mathcal{N}=2$ conformal supergravity, where only dilaton Weyl multiplets have been identified \cite{Adhikari:2023tzi, Adhikari:2024esl}. This is attributed to the absence of a rigid superconformal algebra as dictated by Nahm's classification \cite{Nahm:1977tg}, thereby precluding the existence of a standard Weyl multiplet.}. The primary distinction between these two lies in their covariant matter field content. The dilaton Weyl multiplet is characterized by the presence of a physical scalar field with non-zero Weyl weight. 

The standard procedure to construct a dilaton Weyl multiplet involves coupling on-shell matter multiplets to the standard Weyl multiplet and interpreting the field equations of the matter multiplets as constraints that make some of the standard Weyl multiplet fields composite.  
The dilaton Weyl multiplet has been constructed for dimensions less than or equal to six, accommodating various amounts of supersymmetry \cite{Fujita:2001kv,Bergshoeff:2001hc, Bergshoeff:1985mz, Butter:2017pbp, Hutomo:2022hdi, Gold:2022bdk, Adhikari:2023tzi, Ciceri:2024xxf, Adhikari:2024esl}. The construction of Poincar{\'e} supergravity using dilaton Weyl multiplet instead of the standard Weyl multiplet in the superconformal approach has various advantages. Firstly, since the dilaton Weyl multiplet involves a certain number of compensating matter multiplets coupled to the standard Weyl multiplet by construction, any construction of Poincar{\'e} supergravity using dilaton Weyl multiplet will require fewer compensating multiplets. Secondly, the dilaton Weyl multiplets come with a 2-form gauge field inbuilt and hence the construction of higher derivative supergravity actions using dilaton Weyl multiplets has a natural connection with string theory where such a 2-form gauge field exists. 

Recently, a dilaton Weyl multiplet for $\mathcal{N}=3$ conformal supergravity in four dimensions was constructed by coupling an on-shell vector multiplet to the standard Weyl multiplet  \cite{Adhikari:2024qxg}. In this construction, some of the auxiliary fields of the $\mathcal{N}=3$ standard Weyl multiplet becomes composite while keeping the number of off-shell degrees of freedom intact. In this paper, we perform a similar construction by coupling two on-shell vector multiplets to the standard Weyl multiplet. This yields a new $\mathcal{N}=3$ dilaton Weyl multiplet in four dimensions. In this variant, more number of auxiliary fields of the standard Weyl multiplet becomes composite in contrast to the previous case, while keeping the number of off-shell degrees of freedom intact. As a result of this, the number of compensating vector multiplets required to go from $\mathcal{N}=3$ conformal to Poincar{\'e} supergravity reduces from three to two to one while using the standard \cite{Hegde:2018mxv,vanMuiden:2017qsh}, the old dilaton Weyl multiplet \cite{Adhikari:2024qxg} or the new dilaton Weyl multiplet constructed in this paper.  To avoid confusion with the previously constructed multiplet \cite{Adhikari:2024qxg}, we refer to our result as the ``variant $\mathcal{N}=3$ dilaton Weyl multiplet" in four dimensions.

This paper is organized as follows. Section \ref{N=3} briefly reviews the standard Weyl multiplet and on-shell vector multiplet in $\mathcal{N}=3$ conformal supergravity. In section \ref{Dilatonweyl}, we carry out the construction of the variant $\mathcal{N}=3$ dilaton Weyl multiplet and give the details of its field content and supersymmetry transformations. Finally, in section \ref{conclusions}, we summarize our results and discuss some future directions.  

\section{{$\mathcal{N}=3$} Conformal Supergravity} \label{N=3}

An $\mathcal{N}$ extended conformal supergravity in four dimensions is based on the gauge theory of the superconformal algebra $SU(2,2|\mathcal{N})$,  satisfying certain curvature constraints. 
$\mathcal{N}=2$ and $\mathcal{N}=4$ conformal supergravities have been extensively studied in the past. However, detailed studies on $\mathcal{N}=3$ conformal supergravity began only in the last decade \cite{ vanMuiden:2017qsh, Hegde:2021rte,Hegde:2022wnb, Kuzenko:2023qkg,Adhikari:2024qxg}. The following subsections briefly review the standard Weyl multiplet and the vector multiplet in $\mathcal{N}=3$ conformal supergravity.
 
\subsection{{$\mathcal{N}=3$}\label{stdweyl} Standard Weyl Multiplet }
Robust studies on $\mathcal{N}=3$ conformal supergravity in four dimensions were made possible by the introduction of the $\mathcal{N}=3$ standard Weyl multiplet in the last decade{\cite{vanMuiden:2017qsh,Hegde:2018mxv}}. The details of the independent fields of the standard Weyl multiplet are given in Table $\eqref{standard4d}$. 
\begin{widetext}
\begin{center}
\begin{table}
		\centering
		\begin{tabular}{|P{2cm}|P{5cm}|P{3cm}|P{1cm}|P{1cm}|}
			\hline
			Field&Properties&SU(3) irreps&$w$ & $c$\\
			\hline
			\multicolumn{5}{|c|}{Independent Gauge fields}\\
			\hline
			$e_\mu^a$&vielbein&\bf{1}&$-1$&$0$\\
			$\psi_\mu^I$&$\gamma_5 \psi_\mu^I=\psi_\mu^I$\;, gravitino &\bf{3}&$-1/2$&$-1/2$\\
			$V_\mu{}^{I}{}_J $&$(V_\mu{}^I{}_J)^*\equiv V_{\mu I}{}^J=-V_\mu{}^J{}_I $\;, $SU(3)_R$ gauge field &\bf{8}&0&0\\
			$A_\mu$&$ U(1)_R $ gauge field&\bf{1}&0&0\\
            $b_\mu$&Dilatation gauge field&\bf{1}&0&0\\
			\hline
			\multicolumn{5}{|c|}{Covariant fields}\\
			\hline
 $  T^I_{ab}$& Self-dual i.e $T^I_{ab}=\frac{1}{2}\varepsilon_{abcd}T^{I cd} $&\bf{3}&$1$&$1$\\
 $E_I$& Complex & $\bf{\Bar{3}}$ &$1$&$-1$\\
$ D^I{}_J$& $(D^I{}_J)^*\equiv D_I{}^J=D^J{}_I $&\bf{8}&2&0\\
$\chi_{IJ}$&$\gamma_5\chi_{IJ}=\chi_{IJ} $& $\bf{\Bar{6}}$&$3/2$&$-1/2$\\
$\zeta^I$ & $\gamma_5\zeta^I=\zeta^I $&\bf{3}& $3/2$&$-1/2$\\
$\Lambda_L$&$\gamma_5\Lambda_L=\Lambda_L$&\bf{1}&$1/2$&$-3/2$\\
 
			\hline
		\end{tabular}
		\caption{Field content of the $\mathcal{N}=3$ standard Weyl multiplet}
		\label{standard4d}	
	\end{table}
    \end{center}
\end{widetext}
Conformal supergravity has two kinds of supersymmetries, namely the Q (Ordinary) and the S (Special) supersymmetries, parametrized by the Majorana spinors $\epsilon^I$ and $\eta^I$ respectively. Here $I=1,2,3$ is an $SU(3)$ index. The bosonic symmetries are the covariant general coordinate transformations, Lorentz transformations, $SU(3) \times U(1)$ R symmetry, dilatation, and special conformal transformations. The Weyl weight $(w)$ and chiral weight $(c)$ indicate how the fields transform under dilatation and $U(1)_R$ symmetry, respectively. Specifically, a field X with a Weyl weight $w$ transforms under dilataion as $\delta_D X = w\Lambda_DX$ and a field with a chiral weight $c$ transforms under $U(1)_R$ as $\delta_{U(1)_R} X = ic\Lambda_A X$, where $\Lambda_D$ and $\Lambda_A$ are the parameters of dilatation and $U(1)_R$ respectively. The dilatation gauge field $b_\mu$ is the only independent field in the Weyl multiplet that transforms non-trivially under special conformal transformations (SCT) as $\delta_K b_\mu = \Lambda_{K \mu}$, where $\Lambda_{K \mu}$ is the SCT parameter.  The S-supersymmetry gauge field ($\phi_{\mu}^i$), SCT gauge field $(f_{\mu}{}^a)$ and the spin connection $\omega_{\mu}{}^{ab}$ are composite fields by virtue of the curvature constraints (see the appendices of \cite{Hegde:2022wnb}).
\begin{widetext}    
The Q and S supersymmetry transformations of the components of the standard Weyl multiplet are given below in equations-\eqref{standtrans}:
\allowdisplaybreaks
\begin{subequations}\label{standtrans}
\begin{align}
    \delta e_{\mu}^{a}&= \bar{\epsilon}_{I}\gamma^{a}\psi_{\mu}^{I}+\text{h.c.}\\[5pt]
	\delta \psi_{\mu}^{I}&=2\mathcal{D}_{\mu}\epsilon^{I}-\frac{1}{8}\varepsilon^{IJK}\gamma\cdot T_{J}\gamma_{\mu}\epsilon_{K}-\varepsilon^{IJK}\bar{\epsilon}_{J}\psi_{\mu K}\Lambda_{L}-\gamma_{\mu}\eta^{I} \label{gravitino}\\[5pt]
 \delta V_\mu{}^I{}_J &=\bar{\epsilon}^I\phi_{\mu J}- \frac{1}{48}\bar{\epsilon}^I\gamma_\mu\zeta_J+ \frac{1}{16}\varepsilon_{JKM}\bar{\epsilon}^K\gamma_\mu\chi^{IM}- \frac{1}{16}\bar{\epsilon}^I\gamma\cdot T_J \gamma_\mu\Lambda_R- \frac{1}{16}\bar{\epsilon}^I\gamma_\mu \Lambda_R E_J \nonumber \\
	&+\frac{1}{8}\varepsilon_{KMJ}E^I\bar{\epsilon}^K\psi_\mu^M  + \frac{1}{4}\bar{\epsilon}^I\gamma^a\psi_{\mu J}\bar{\Lambda}_L\gamma_a\Lambda_R-\bar{\psi}_\mu^I\eta_J-\text{h.c.}-\text{trace} \\[5pt]
 \delta A_\mu &=\frac{i}{6}\bar{\epsilon}^I\phi_{\mu I}+ \frac{i}{36}\bar{\epsilon}^I\gamma_\mu\zeta_I+ \frac{i}{12}\varepsilon_{KMP}E^P\bar{\epsilon}^K\psi_{\mu}^M+ \frac{i}{12}\bar{\epsilon}^I\gamma\cdot T_I\gamma_\mu\Lambda_R+\frac{i}{12}\bar{\epsilon}^I\gamma_\mu\Lambda_RE_I\nonumber\\
	&-\frac{i}{3}\bar{\epsilon}^I\gamma^a\psi_{\mu I}\bar{\Lambda}_L\gamma_a\Lambda_R-\frac{i}{6}\bar{\psi}_\mu^I\eta_I+\text{h.c.
}\\[5pt]
\delta b_\mu &= \frac{1}{2}(\bar{\epsilon}^I\phi_{\mu I}-\bar{\psi}_\mu^I\eta_I)+\text{h.c.}\\[5pt]
	\delta \Lambda_L&=-\frac{1}{4}E_I\epsilon^I+\frac{1}{4}\gamma\cdot T_I\epsilon^I\\[5pt]
	\delta E_I &=-4 \bar{\epsilon}_I\cancel{D}\Lambda_L-\frac{1}{2}\varepsilon_{IJK}\bar{\epsilon}^J\zeta^K+\frac{1}{2}\bar{\epsilon}^J\chi_{IJ}-\frac{1}{2}\varepsilon_{IJK}E^K\bar{\epsilon}^J\Lambda_L-4\bar{\Lambda}_L\Lambda_L\bar{\epsilon}_I\Lambda_R- 4\bar{\eta}_I\Lambda_L \\[5pt]
	\delta T^I_{ab} &= -\bar{\epsilon}^I\cancel{D}\gamma_{ab}\Lambda_R-4\varepsilon^{IJK}\bar{\epsilon}_JR_{ab}(Q)_K+\frac{1}{8}\bar{\epsilon}_J\gamma_{ab}\chi^{IJ}+\frac{1}{24}\varepsilon^{IJK}\bar{\epsilon}_J\gamma_{ab}\zeta_K \nonumber \\
	&-\frac{1}{8}\varepsilon^{IJK}E_J\bar{\epsilon}_K\gamma_{ab}\Lambda_R+\bar{\eta}^I\gamma_{ab}\Lambda_R \\[5pt]
	\delta \chi_{IJ}&=2\cancel{D}E_{(I}\epsilon_{J)}-8\varepsilon_{KM(I}\gamma\cdot R(V)^M{}_{J)}\epsilon^K-2\gamma\cdot\cancel{D}T_{(I}\epsilon_{J)}+\frac{1}{3}\varepsilon_{KM(I}D^M{}_{J)}\epsilon^K\nonumber \\
 &+\frac{1}{4}\varepsilon_{KM(I}E^K\gamma\cdot T_{J)}\epsilon^M-\frac{1}{3}\bar{\Lambda
	}_L\gamma_a\epsilon_{(I}\gamma^a\zeta_{J)}+\frac{1}{4}\varepsilon_{PM(I}E_{J)}E^M\epsilon^P-\bar{\Lambda}_L\gamma^a\Lambda_R\gamma_aE_{(I}\epsilon_{J)}\nonumber\\
	&-\bar{\Lambda}_L\gamma\cdot T_{(I}\gamma^a\Lambda_R\gamma_a\epsilon_{J)}+ 2\gamma\cdot T_{(I}\eta_{J)}+ 2E_{(I}\eta_{J)}\\[5pt]
	\delta \zeta^I &=- 3\varepsilon^{IJK}\cancel{D}E_J\epsilon_K +\varepsilon^{IJK}\gamma\cdot\cancel{D}T_K\epsilon_J-4\gamma\cdot R(V)^I{}_J\epsilon^J-16i\gamma\cdot R(A)\epsilon^I-\frac{1}{2}D^I{}_J\epsilon^J\nonumber\\
	&-\frac{3}{8}E^I\gamma\cdot T_J\epsilon^J+\frac{3}{8}E^J\gamma\cdot T_J\epsilon^I+\frac{3}{8}E^IE_J\epsilon^J+\frac{1}{8}E^JE_J\epsilon^I- 4 \bar{\Lambda}_L\cancel{D}\Lambda_{R}\epsilon^I- 4 \bar{\Lambda}_R\cancel{D}\Lambda_L\epsilon^I\nonumber\\
	&- 3\bar{\Lambda}_R\cancel{D}\gamma_{ab}\Lambda_L\gamma^{ab}\epsilon^I-3\bar{\Lambda}_L\gamma_{ab}\cancel{D}\Lambda_R\gamma^{ab}\epsilon^I+\frac{1}{2}\varepsilon^{IJK}\bar{\Lambda}_L\gamma^a\epsilon_J\gamma_a\zeta_K-6\bar{\Lambda}_L\Lambda_L\bar{\Lambda}_R\Lambda_R\epsilon^I\nonumber\\
	&+\varepsilon^{IJK}\gamma\cdot T_J\eta_K-3\varepsilon^{IJK}E_J\eta_K\\[5pt]
	\delta D^I{}_J&=-3\bar{\epsilon}^I\cancel{D}\zeta_J-3\varepsilon_{JKM}\bar{\epsilon}^K\cancel{D}\chi^{IM}+\frac{1}{4}\varepsilon_{JKM}\bar{\epsilon}^I\zeta^K E^M +\frac{1}{2}\varepsilon_{JKM}\bar{\epsilon}^K\zeta^M E^I +\frac{3}{4}\bar{\epsilon}^I\chi_{JK}E^K\nonumber\\
	&+ 3\bar{\epsilon}^I\gamma\cdot T_J\overset{\leftrightarrow}{\cancel{D}}\Lambda_R-3\bar{\epsilon}^I\cancel{D}\Lambda_RE_J-3\bar{\epsilon}^I\cancel{D}E_J\Lambda_R+ \frac{3}{4}\varepsilon_{JKM}E^M\bar{\epsilon}^K\Lambda_LE^I\nonumber\\
&+ {3\varepsilon_{JKM}T^I\cdot T^M\bar{\epsilon}^K\Lambda_{L}}-2\bar{\epsilon}^I\Lambda_L\bar{\Lambda}_R\zeta_J-3\bar{\epsilon}^I\Lambda_L\bar{\Lambda}_R\Lambda_RE_J+3\bar{\epsilon}^I\gamma\cdot T_J\Lambda_L\bar{\Lambda}_R\Lambda_R\nonumber\\
&+\text{h.c} -\text{trace}
	\end{align}
    \end{subequations}
\end{widetext}

  \textbf{Notation:} 
To be consistent with the existing literature on four dimensional conformal supergravity, we follow the chiral notation here. In this notation, raising and lowering of $SU(3)$ R-symmetry indices is done via complex conjugation. For example, for complex fields such as $E^I$ in the $\NN=3$ standard Weyl multiplet, the relation is $E_I=(E^I)^*$. For real fields, raising and lowering of the R-symmetry indices is also accompanied by changing the duality or chirality property of the fields. For example, we denote the anti-self dual part of a real anti-symmetric tensor field in the $\NN=3$ standard Weyl multiplet as $T_{abI}$. Under complex conjugation, it goes to $T_{ab}^{I}$ which is the self-dual part of the same anti-symmetric tensor field. In the same way the left and right chiral components of a Majorana spinor are denoted with different positioning of the R-symmetry indices and are related to each other via complex conjugation. For example, the left and right chiral component of a Majorana spinor in the standard Weyl multiplet are denoted as $\zeta^I$ and $\zeta_I$ respectively, which are related to each other as $\zeta_I=i\gamma^0 C^{-1}(\zeta^I)^*$, where $C$ is the charge conjugation matrix. There is a Majorana fermionic field $\Lambda$ in $\NN=3$ standard Weyl mutliplet which does not carry any R-symmetry indices. In that case, the left and right chiral components also satisfy similar relation i.e $\Lambda_R=i\gamma^0 C^{-1}(\Lambda_L)^*$, where the subscript $R$ and $L$ denote the left and right chiral part of the spinor and not any R-symmetry index. 

The notation $D_a$ is used to denote a fully supercovariant derivative. $\mathcal{D}_{\mu}\epsilon^I$ appearing in the supersymmetry transformations of the gravitino \eqref{gravitino} is defined as:
 \begin{align}
     \mathcal{D}_{\mu}\epsilon^I=\partial_\mu \epsilon^I-\frac{1}{4}\gamma\cdot
      \omega_\mu \epsilon^I+\frac{1}{2}(b_\mu+iA_\mu)\epsilon^I-V_\mu{}^I{}_J \epsilon^J
 \end{align}

\subsection{{$\mathcal{N}=3$} Vector Multiplets} \label{vector}
The $\mathcal{N}=3$ vector multiplet coupled to the standard Weyl multiplet was constructed in \cite{Hegde:2022wnb}. The field contents are given in Table \eqref{4dvector}. In this paper, we consider two vector multiplets coupled to conformal supergravity. To distinguish the two vector multiplets, we label them with indices $\mathcal{A, B..} \in \{1,2 \}$.  
\begin{table}[h!]
	\centering
	\begin{tabular}{ |P{2cm}|P{2.5cm}|P{1cm}|P{1cm}|P{1cm}|}
		\hline
		Field & Type& SU(3)&$w$&$c$\\
		\hline
		$C_\mu{}^{\mathcal{A}}$& Boson&\bf{1}&0&0\\
		$\xi_I{}^{\mathcal{A}}$&Boson& $\bar{ \bf{3}}$ &$1$&$-1$\\
		$\psi_{I}{}^{\mathcal{A}}$&{Fermion}&$ \bar{\bf{{3}}}$&$3/2$&$1/2$\\
  $\theta_L^{\mathcal{A}}$ & Fermion & \bf{1}& $3/2$& $3/2$\\
		\hline
	\end{tabular}
	\caption{The field content of $\mathcal{N}=3$ on-shell vector multiplets}
	\label{4dvector}	
\end{table}
\begin{widetext}
The Q and S transformations of the components of this multiplet are given below in equations-\eqref{vec}.
\allowdisplaybreaks
{\begin{subequations}\label{vec}
\begin{align}
\delta{C}_{\mu}{}^{\mathcal{A}}&= \bar{\epsilon}^{I}\gamma_{\mu}\psi_{I}^{\mathcal{A}}-2\varepsilon_{IJK}\bar{\epsilon}^{I}\psi_{\mu}^J\xi^{K \mathcal{A}}  - \bar{\epsilon}_{I}\gamma_{\mu}\Lambda_L \xi^{I \mathcal{A}} +\text{h.c.}    \\
\delta \psi_{I}{}^{\mathcal{A}}&= -\frac{1}{2}   \gamma \cdot \mathcal{F}^{+ \mathcal{A}} \epsilon_I - 2 \varepsilon_{IJK} \cancel{D} \xi^{K \mathcal{A}} \epsilon^J - \frac{1}{4} E_I \xi^{J \mathcal{A}} \epsilon_J + \frac{1}{2} \bar{\Lambda}_L \theta_L^{\mathcal{A}} \epsilon_I\nonumber \\
& + \frac{1}{2}\gamma_a \epsilon^J \bar{\Lambda}_R \gamma^a \Lambda_L \xi^{K \mathcal{A}} \varepsilon_{IJK} + 2 \varepsilon_{IJK} \xi^{J \mathcal{A}} \eta^K 
   \\
\delta \theta_L^{\mathcal{A}} &= - 2 \cancel{D} \xi^{I \mathcal{A}} \epsilon_I -  \gamma^a \bar{\Lambda}_L \gamma_a \Lambda_R \xi^{I \mathcal{A}} \epsilon_I + \frac{1}{4} \varepsilon_{IJK} E^I \xi^{J \mathcal{A}} \epsilon^K - \bar{\Lambda}_R \psi_I^{\mathcal{A}} \epsilon^I  - 2 \xi^{I \mathcal{A}} \eta_I 
   \\
\delta \xi_I{}^{\mathcal{A}}&=-\bar{\epsilon_I} \theta_R^{\mathcal{A}} + \varepsilon_{IJK} \bar{\epsilon}^J \psi^{K \mathcal{A}}  
\end{align}    
\end{subequations}}
\end{widetext}

where $\mathcal{F}_{ab}^{+ \mathcal{A}}$ is the self-dual part of a  modified super-covariant field strength of $C_\mu^{\mathcal{A}}$, given as,
\begin{align}
 \mathcal{F}_{a b}^{+\mathcal{A}}= &{F}_{a b}^{+ \mathcal{A}}-\frac{1}{4} \bar{\Lambda}_R \gamma_{a b} \theta_R^{\mathcal{A}}-\frac{1}{2} T_{a b}^I\xi_I{}^{\mathcal{A}},
 \end{align}

 where,
 \begin{align}
   {F}_{ab}{}^{\mathcal{A}} =  2 e_{[a}^\mu e_{b]}^\nu \partial_\mu C_\nu^{\mathcal{A}} - &\left( \bar{\psi}^I_{[a} \gamma_{b]} \psi_I^{\mathcal{A}} - \epsilon_{IJK} \bar{\psi}_{[a}^I \psi_{b]}^{J} \xi^{K \mathcal{A}}\right.\nonumber \\
   & \left.- \bar{\psi}_I{}_{[a} \gamma_{b]} \Lambda_L \xi^{I \mathcal{A}} + h.c\right)
\end{align}
is the fully super-covariant field strength associated with the vector gauge field $C_\mu{}^\mathcal{A}$. The $\mathcal{N}=3$  vector multiplets are on-shell multiplets. Hence the superconformal algebra closes on its fields only when their field equations are satisfied. These field equations play a crucial role in the following sections of this paper. They are given below,
\begin{widetext}
\allowdisplaybreaks{
\begin{subequations} \label{eom}
    \begin{align}
& \cancel{D} \psi_I{}^{\mathcal{A}}+\frac{1}{2} \bar{\Lambda}_R \psi_I^{\mathcal{A}} \Lambda_L-\frac{1}{8} E_I \theta_L^{\mathcal{A}}+\frac{1}{8} \gamma \cdot T_I \theta_L^{\mathcal{A}}+\frac{1}{8} \chi_{I J} \xi^{J \mathcal{A}}+\frac{1}{24} \varepsilon_{IJK} \zeta^J \xi^{K \mathcal{A}}=0\;,\label{psi} \\[5pt]
& \cancel{D} \theta_R^{\mathcal{A}}-\frac{3}{4} \bar{\Lambda}_R \theta_R^{\mathcal{A}} \Lambda_L+\frac{1}{4} \gamma \cdot F^{- \mathcal{A}} \Lambda_L-\frac{3}{8} \bar{\Lambda}_L \Lambda_L \theta_L^{\mathcal{A}}-\frac{1}{8} \gamma \cdot T_I \Lambda_L \xi^{I \mathcal{A}}-\frac{1}{8} E_I \psi^{I \mathcal{A}}-\frac{1}{8} \gamma \cdot T_I \psi^{I \mathcal{A}} \nonumber \\
& -\frac{1}{12} \zeta^I \xi_I^{\mathcal{A}}-\frac{1}{8} E^I \xi_I^{\mathcal{A}} \Lambda_L=0 \;,\label{theta}\\[5pt]
&\square_c \xi_J{}^{\mathcal{A}}+ \frac{1}{4} D_a (\bar{\Lambda}_R \gamma_a \Lambda_L \xi_J{}^{\mathcal{A}} ) + \frac{1}{4} \bar{\Lambda}_R \cancel{D} \xi_J{}^{\mathcal{A}} \Lambda_L+\frac{1}{4} F^{- \mathcal{A}} \cdot T_J-\frac{1}{16} \bar{\Lambda}_L \gamma \cdot T_J \theta_L^{\mathcal{A}}-\frac{1}{8} \xi^{I \mathcal{A}} T_I \cdot T_J\nonumber \\
&+\frac{1}{24} \bar{\zeta}_J \theta_R^{\mathcal{A}} +\frac{1}{16} E_J \bar{\Lambda}_R \theta_R^{\mathcal{A}}  -\frac{1}{16} \bar{\chi}_{M J} \psi^{M \mathcal{A}}-\frac{1}{48} \varepsilon_{J K M} \bar{\zeta}^M \psi^{K \mathcal{A}}-\frac{1}{48} D^I{ }_J \xi_{I}{}^\mathcal{A}-\frac{1}{96} \xi_{J}{}^\mathcal{A} E^K E_K\nonumber \\
&+\frac{1}{12} \xi_{J} ^\mathcal{A}\left(\bar{\Lambda}_R \cancel{D} \Lambda_L+\bar{\Lambda}_L \cancel{D} \Lambda_R\right)  +\frac{1}{12} \xi_{J} ^\mathcal{A} \bar{\Lambda}_R \Lambda_R \bar{\Lambda}_L \Lambda_L=0\;,\label{xi}\\[5pt]
&D_a\left({G}^{+ \mathcal{A}}_{a b}-{G}^{- \mathcal{A}}_{a b}\right)=0\;,\label{maxwell}
\end{align}
\end{subequations}
}
\end{widetext}
where  $\square_c$ is the superconformal de-Alembertian. The field strength $G_{ab}^{\mathcal{A}}$ appearing in the Maxwell's equations \eqref{maxwell}is defined as,
 \begin{align} \label{G}
       {G}_{ab}^{+ \mathcal{A}}=& - i F_{ab}^{+ \mathcal{A}} +\frac{i}{2} \bar{\Lambda}_R \gamma_{ab} \theta_R^{\mathcal{A}} + iT_{ab}^I \xi_I^{\mathcal{A}} \; .
 \end{align}
 \section{Variant dilaton Weyl multiplet with R symmetry \texorpdfstring{$SU(2)\times U(1) \times U(1) $}{SU(2)X U(1)X U(1)}} \label{Dilatonweyl}

 Recently in \cite{Adhikari:2024qxg}, we constructed a dilaton Weyl multiplet in $\mathcal{N}=3$ conformal supergravity by coupling one on-shell vector multiplet to the standard Weyl multiplet. In this section, we discuss the construction of a variant dilaton Weyl multiplet by coupling two on-shell vector multiplets to the standard Weyl multiplet. The procedures we follow in this construction are more or less the same as those used in \cite{Adhikari:2024qxg}. We use the field equations of the
vector multiplets to solve for some of the components of the auxiliary fields in the standard
Weyl multiplet in terms of the fields from the vector multiplet and some newly introduced dual gauge
fields.\footnote{Dual gauge fields appear when we interpret some field equations as the Bianchi identities of these dual gauge fields, as we will see later in this section.} The remaining independent fields of the standard Weyl multiplet along with the fields of the two vector multiplets and the dual gauge fields constitute a new off-shell multiplet, which we refer to as the variant $\mathcal{N}=3$ dilaton Weyl multiplet. 

Before trying to solve the field equations, we first need to break the $SU(3)$ R- symmetry to its subgroup $SU(2) \times U(1)$. This breaking plays a crucial role in allowing us to use the field equations to solve some of the Weyl multiplet fields algebraically. Under this breaking, the fields and symmetry parameters that previously formed irreducible representations of $SU(3)$ will decompose into irreducible representations of $SU(2) \times U(1).$ These decompositions are given in appendix \ref{decomposition}. The $SU(3)$ indices $I,J\cdots=1,2,3$ decomposes to $SU(2)$ indices $i,j\cdots=1,2$ and $I=3$. The relation between the $SU(3)$ Levi-Civita and $SU(2)$ Levi Civita is as follows:
\begin{align}
    \varepsilon_{ij3}&=\varepsilon_{ij} \;.
\end{align}
Under this decomposition, we denote some fields with a mathring (eg: $\mathring{\chi_i}, \mathring{\chi_{ij}}$, $\mathring{\zeta}_L$ ...) anticipating that they will become composite while solving the field equations, as we will see later in this section.

The scalar fields $\xi_I{}^\mathcal{A}$ will decompose into $\xi_i{}^\mathcal{A}, \xi_3{}^\mathcal{A}$. 
We gauge fix  $SU(3)$ to $SU(2)\times U(1)$ by setting:
\begin{align} \label{gauge fixing}
    \xi_3{}^\mathcal{A}=0 \;.
\end{align}
The original Q-supersymmetry transformation does not preserve the above gauge fixing choice and hence an unbroken $Q$-supersymmetry needs to be defined in the following way by adding a compensating $SU(3)$ transformation. 
\begin{align} \label{compensate}
\delta_Q^{\text{new}}(\epsilon^i,\epsilon_L)= \delta_Q(\epsilon)+\delta_{SU(3)}(\Lambda^i{}_3=u(\epsilon)^i) \;,   
\end{align}
where,
\begin{align}\label{udef}
    u(\epsilon)^i&=\bigg(-\bar{\epsilon}_R\theta_R^\mathcal{A}+\varepsilon_{jk}\bar{\epsilon}^j\psi^{k\mathcal{A}} \bigg)\xi_\mathcal{A}{}^i \;.
\end{align}
From here on, only $\delta_Q^{\text{new}}$ will appear and hence we drop the superscript ``new" for convenience. We will consider $\xi_i{}^\mathcal{A}$ as a $2\times 2$ matrix-valued complex field. The inverse of this field is denoted as $\xi_{\mathcal{A}}{}^i$, which appears in \eqref{udef}. The complex conjugate of the inverse according to the chiral notation is denoted as $\xi_{\mathcal{A}i}$. Using the same chiral notation, the complex conjugate of $\xi_{i}{}^{\mathcal{A}}$ denoted as $\xi^{i\mathcal{A}}$. By construction $\xi_{\mathcal{A}i}$ is also the inverse of $\xi^{i\mathcal{A}}$. To summarize, the following relations hold:
\begin{align}
    &\xi_{i}{}^{\AAA}\xi_{\AAA}{}^j=\delta_i{}^j\;,\;\; \xi_{\AAA}{}^i\xi_{i}{}^{\BB}=\delta_{\AAA}{}^{\BB}\;,\nonumber \\
    &\xi^{i\AAA}=(\xi_{i}{}^{\AAA})^*\;, \;\; \xi_{\AAA i}=(\xi_{\AAA}{}^{i})^*\;, \nonumber \\
    & \xi^{i\AAA}\xi_{\AAA j}=\delta^i{}_j\;, \;\; \xi_{\AAA i}\xi^{i\BB}=\delta_{\AAA}{}^{\BB}\;.
\end{align}

Let us make a few comments on the decomposition of the $SU(3)$ gauge field $V_\mu{}^{I}{}_{J}$ under the gauge fixing. The following combinations of the decomposed fields becomes the $SU(2)$ gauge field $\mathring{v}_\mu{}^i{}_j$ and $U(1)$ gauge field $\mathring{v}_\mu$. 
\begin{align}
    \mathring{v}_\mu{}^i{}_j &\equiv V_\mu{}^i{}_j - \frac{1}{2} \delta^i_j V_\mu{}^3{}_3 \nonumber \\
    \mathring{v}_\mu &\equiv -i V_\mu{}^3{}_3
\end{align}
We now have two $U(1)$ R-symmetries in the theory, one coming from the breaking of the $SU(3)$ R-symmetry and the other which is already present in $\NN=3$ conformal supergravity. To distinguish them we refer to the former as $U(1)_v$ and the latter as $U(1)_A$. The symmetry parameters are denoted as $\lambda_v$ and  $\lambda_A$, respectively. The weights of any field w.r.t these $U(1)$ symmetries are denoted as $c_v$ and $c_A$ respectively. 

The $V_a{}^i{}_3$ component of the $SU(3)$ gauge field combines with gravitino-dependent terms, as shown in \eqref{Ydef} to form a covariant matter field, which we denote as $Y_a^i$.

\begin{align}\label{Ydef}
    Y_a^i \equiv V_a{}^i{}_3 - \frac{1}{2} u(\psi_a)^i
\end{align}

The supercovariant derivatives $D_a$ that appeared in the previous sections are supercovariant w.r.t. the old supersymmetry transformations and the $SU(3)\times U(1)$ R-symmetry of the standard Weyl multiplet, along with the other gauge symmetries in theory. After gauge fixing, they will decompose into a new supercovariant derivative plus some supercovariant terms containing $Y_a^i$, where the new supercovariant derivative is covariant w.r.t. to the new supersymmetry transformations as well $SU(2)\times U(1)\times U(1)$ R-symmetry of the dilaton Weyl multiplet. For example, the supercovariant derivative of the gaugino $\psi^i$ decomposes as follows, 

\begin{align}
    D_a^{\text{old}}\psi^i = D_a^{\text{new}} \psi^i - Y_a^i \psi_R
\end{align}

Only $D_a^{new}$ will appear in what follows, and we drop the superscript ``new" for convenience.

Now we proceed to analyze the field equations \eqref{eom}. Let us write them in terms of the new supercovariant derivative and fields in $SU(2)\times U(1)$ representations.
\begin{widetext}
\allowdisplaybreaks
\begin{subequations}
\begin{align}
    &\cancel{D}\psi_i{}^\mathcal{A}-\slashed{Y}_i \psi_R^\mathcal{A} + \frac{1}{2} \bar{\Lambda}_R \psi_i{}^\mathcal{A} \Lambda_L - \frac{1}{8} E_i \theta_L^\mathcal{A} + \frac{1}{8} \gamma \cdot \mathring{T}_i \theta_L^\mathcal{A}+\frac{1}{8} \mathring{\chi}_{ij} \xi^{j \mathcal{A}} - \frac{1}{24} \varepsilon_{ij} \mathring{\zeta}_L \xi^{j \mathcal{A}} =0\\[5pt]
  & \cancel{D}\psi_R^\mathcal{A} - \slashed{Y}^i \psi_i{}^\mathcal{A} + \frac{1}{2} \bar{\Lambda}_R \psi_R^\mathcal{A} \Lambda_L - \frac{1}{8} E \theta_L^\mathcal{A} + \frac{1}{8} \gamma \cdot T^- \theta_L^\mathcal{A} + \frac{1}{8} \mathring{\chi}_j \xi^{j\mathcal{A}} + \frac{1}{24} \varepsilon_{ij} \mathring{\zeta}^i \xi^j{^\mathcal{A}} = 0 \\[5pt]
  &  \cancel{D}\theta_R^\mathcal{A}- \frac{3}{4} \bar{\Lambda}_R \theta_R^\mathcal{A} \Lambda_L + \frac{1}{4} \gamma \cdot F^{- \mathcal{A}} \Lambda_L - \frac{3}{8} \bar{\Lambda}_L \Lambda_L \theta_L^\mathcal{A} - \frac{1}{8} \gamma \cdot \mathring{T}_i \Lambda_L \xi^{i \mathcal{A}} - \frac{1}{8} \gamma \cdot T^- \psi_L^\mathcal{A}   - \frac{1}{12} \mathring{\zeta}^i \xi_i{}^\mathcal{A} \nonumber \\
    &\;- \frac{1}{8} E^i \xi_i{}^\mathcal{A} \Lambda_L \Lambda_L =0
  \\[5pt]
  &  \Box \xi_i{}^\mathcal{A}+ Y^a_i Y_a{}^j \xi_j{}^{\mathcal{A}} + \frac{1}{4} D_a (\bar{\Lambda}_R \gamma_a \Lambda_L \xi_i{}^{\mathcal{A}} ) + \frac{1}{4} \bar{\Lambda}_R \cancel{D} \xi_i{}^{\mathcal{A}} \Lambda_L+ \frac{1}{4} \widehat{F}^{- \mathcal{A}} \cdot \mathring{T}_i - \frac{1}{16} \bar{\Lambda_L} \gamma \cdot \mathring{T}_i \theta_L^{\mathcal{A}}  - \frac{1}{8} \xi^{j \mathcal{A}} \mathring{T}_j \cdot \mathring{T}_i \nonumber \\
   &\;+ \frac{1}{24} \mathring{\bar{\zeta}}_i \theta_R ^\mathcal{A} + \frac{1}{16} E_i \bar{\Lambda}_R \theta_R^\mathcal{A} -\frac{1}{16} \mathring{\bar{\chi}}_{ij} \psi^{j \mathcal{A}} - \frac{1}{16} \mathring{\bar{\chi}}_i \psi_L^\mathcal{A} - \frac{1}{48} \varepsilon_{ij} \mathring{\bar{\zeta}}_L \psi^{j \mathcal{A}}  - \frac{1}{48} \mathring{D}^j{}_{i} \xi_j^\mathcal{A} + \frac{1}{96} \mathring{D} \xi_i{}^{\mathcal{A}} \nonumber \\ \; &\; + \frac{1}{48} \varepsilon_{ij} \mathring{\zeta} ^j \psi_L^{ \mathcal{A}}- \frac{1}{96} \xi_i^\mathcal{A} E^j E_j  - \frac{1}{96} \xi_i{}^\mathcal{A} \lvert E \rvert ^2 + \frac{1}{12} \xi_i{}^\mathcal{A} (\bar{\Lambda}_R \cancel{D}\Lambda_L + \bar{\Lambda}_L \cancel{D}\Lambda_R) + \frac{1}{12} \xi_i {}^\mathcal{A} \bar{\Lambda}_R \Lambda_R \bar{\Lambda}_L \bar{\Lambda}_L=0 
  \\[5pt]
   & D^a (Y_a^i \xi_i^{\mathcal{A}}) + \frac{1}{2} \bar{\Lambda}_R \slashed{Y}^i \xi_i^{\mathcal{A}} \Lambda_L + \frac{1}{4} F^{- \mathcal{A}} \cdot T - \frac{1}{16} \bar{\Lambda}_L \gamma \cdot T^- \theta_L^\mathcal{A} - \frac{1}{8} \xi^{i \mathcal{A}} \mathring{T}_i \cdot T^- + \frac{1}{24} \mathring{\bar{\zeta}}_R \theta_R^{\mathcal{A}}  \nonumber \\ &\;+ \frac{1}{16} E \bar{\Lambda}_R \theta_R^\mathcal{A}- \frac{1}{16} \mathring{\bar{\chi}}_i \psi^{i \mathcal{A}} - \frac{1}{16} \mathring{\bar{\chi}}_L \psi_L^\mathcal{A} - \frac{1}{48} \varepsilon_{ij} \mathring{\zeta}^j \psi^{i \mathcal{A}} - \frac{1}{48} \mathring{D}^i \xi_i^\mathcal{A} =0\\[5pt]
&  D_a\left({G}^{+\mathcal{A}}_{ a b}-{G}^\mathcal{-A}_{ a b}\right)=0\label{newMax}
\end{align}
\end{subequations}
\end{widetext}
\begin{widetext}
By observing the field equations, we can see that the fields $\mathring{D}^i{}_j\;, \mathring{D}^i\;, \mathring{D}\;, \mathring{\chi}_{ij}\;,\mathring{\chi}_i\;,\mathring{\zeta}^i\;,\mathring{\zeta}_L $ can be solved algebraically. By solving the fermionic equations, we get the following relations:

\allowdisplaybreaks

\begin{align}
     \mathring{\chi}_{ij} & = -8 \bigg(\cancel{D}\psi_{(i}{}^\mathcal{A}-\slashed{Y}_{(i} \psi_R^\mathcal{A} + \frac{1}{2} \bar{\Lambda}_R \psi_{(i}^\mathcal{A} \Lambda_L - \frac{1}{8} E_{(i} \theta_L^\mathcal{A} + \frac{1}{8} \gamma \cdot \mathring{T}_{(i} \theta_L^\mathcal{A} \bigg) \xi_{\mathcal{A} j)} \nonumber\\
     \mathring{\zeta}_L &= 12 \varepsilon^{ij} \bigg( \cancel{D}\psi_i{}^\mathcal{A}-\slashed{Y}_i \psi_R^\mathcal{A} + \frac{1}{2} \bar{\Lambda}_R \psi_i^\mathcal{A} \Lambda_L - \frac{1}{8} E_i \theta_L^\mathcal{A} + \frac{1}{8} \gamma \cdot \mathring{T}_i \theta_L^\mathcal{A} \bigg) \xi_{\mathcal{A }j} \nonumber \nonumber\\ 
     \mathring{\chi}_i &= -8\bigg( \cancel{D}\psi_R^\mathcal{A} + \slashed{Y}^i \psi_i^\mathcal{A} + \frac{1}{2} \bar{\Lambda}_R \psi_R^\mathcal{A} \Lambda_L - \frac{1}{8} E \theta_L^\mathcal{A} + \frac{1}{8} \gamma \cdot T^- \theta_L^\mathcal{A}  + \frac{1}{24} \varepsilon_{ij} \mathring{\zeta}^i \xi^j{^\mathcal{A}} \bigg) \xi_{\mathcal{A} i}  \nonumber\\ \mathring{\zeta}^i &= 12\bigg(\cancel{D}\theta_R^\mathcal{A}- \frac{3}{4} \bar{\Lambda}_R \theta_R^\mathcal{A} \Lambda_L + \frac{1}{4} \gamma \cdot F^{- \mathcal{A}} \Lambda_L - \frac{3}{8} \bar{\Lambda}_L \Lambda_L \theta_L^\mathcal{A} - \frac{1}{8} \gamma \cdot \mathring{T}_j \Lambda_L \xi^{j \mathcal{A}} - \frac{1}{8} \gamma \cdot T^- \psi_L^\mathcal{A} \nonumber \\ & -\frac{1}{8} \gamma \cdot \mathring{T}_k \psi^{k \mathcal{A}} -\frac{1}{8} E \psi_L^{\mathcal{A}} - \frac{1}{8} E_i \psi^{i \mathcal{A}} - \frac{1}{8} E^j \xi_j{}^\mathcal{A} \bar{\Lambda}_L \Lambda_L \bigg) \xi_{\mathcal{A}}{}^i \nonumber
\end{align} 
\end{widetext}
 Now let us solve the bosonic equations. Let us consider the Maxwell's equation corresponding to the $U(1)$ gauge fields $C_\mu{}^{\mathcal{A}}$ \eqref{newMax}. We denote $G_{ab}{}^{\AAA}$ defined in \eqref{G} as the field strength of a new pair of $U(1)$ gauge fields $\Tilde{C}_\mu{}^{\mathcal{A}}$. Then the Maxwell's equations \eqref{newMax} becomes the Bianchi identity of this new pair of $U(1)$ gauge fields. These new pair of $U(1)$ gauge fields can be interpreted as the dual of the vector gauge fields $C_{\mu}{}^{\AAA}$. However, now these dual gauge fields become a part of the multiplet as independent gauge fields.

From the definition of $G_{ab}{}^{\AAA}$ \eqref{G}, we can find its transformations. The knowledge of the supersymmetry transformations of $G_{ab}{}^{\mathcal{A}}$, its Bianchi identity \eqref{newMax} and the superconformal algebra are sufficient to obtain the gauge transformations of the dual gauge fields $\Tilde{C}_\mu{}^{\mathcal{A}}$. The result is,
\begin{align} \label{tildeC}
    \delta \Tilde{C}_\mu{}^\mathcal{A}= \partial_{\mu}\Tilde{\lambda}^{\AAA}+&\Bigg[2i(\bar{\epsilon}_L\psi_\mu^j-\bar{\epsilon}^j\psi_{\mu L} )\xi^{k\mathcal{A}}\varepsilon_{jk}+i\bar{\epsilon}_i\gamma_\mu\Lambda_L\xi^{i\mathcal{A}}\nonumber \\
    &-i\bar{\epsilon}^i\gamma_\mu\psi_i{}^\mathcal{A}-i\bar{\epsilon}_L\gamma_\mu\psi_R^{\mathcal{A}}+\text{h.c}\Bigg] \;,
\end{align}
where $\Tilde{\lambda}^{\AAA}$ is the gauge transformation parameter corresponding to the dual gauge fields $\Tilde{C}_\mu{}^{\mathcal{A}}$. 

Now we reinterpret \eqref{G} as a constraint that completely determines $\mathring{T}_{ab}^i$ in terms of the field strengths $F_{ab}{}^{\AAA}$ and $G_{ab}{}^{\AAA}$ as shown below, 

\begin{align}
    \mathring{T}_{ab}^i = F_{ab}^{+ \mathcal{A}} \xi_{}{\mathcal{A}}{}^i - i G_{ab}^{+ \mathcal{A}} \xi_{\mathcal{A}}{}^i - \frac{1}{2} \bar{\Lambda}_R \gamma_{ab} \theta_R^{\mathcal{A}} \xi_{\mathcal{A}}{}^i \;.
\end{align}
\begin{widetext}
The field equation of $\xi^\mathcal{A}$ allows us to make $\mathring{D}^i$ as a composite field, as shown below.
\begin{align}
\mathring{D}^i &= 48 \bigg( D^a (Y_a^j \xi_j{}^{\mathcal{A}}) + \frac{1}{2} \bar{\Lambda}_R \cancel{Y}^j \xi_j{}^{\mathcal{A}} \Lambda_L + \frac{1}{4} F^{- \mathcal{A}} \cdot T - \frac{1}{16} \bar{\Lambda}_L \gamma \cdot T^- \theta_L^\mathcal{A} - \frac{1}{8} \xi^{j \mathcal{A}} \mathring{T}_j \cdot T^-  \nonumber \\ &+ \frac{1}{24} \mathring{\bar{\zeta}}_R \theta_R^{\mathcal{A}} + \frac{1}{16} E \bar{\Lambda}_R \theta_R^\mathcal{A}- \frac{1}{16} \mathring{\bar{\chi}}_j \psi^{j \mathcal{A}} - \frac{1}{16} \mathring{\bar{\chi}}_L \psi_L^\mathcal{A} - \frac{1}{48} \varepsilon_{jk} \mathring{\zeta}^k \psi^{j \mathcal{A}} \bigg) \xi_\mathcal{A}{}^i \;.
\end{align}
From the field equation of $\xi_i^\mathcal{A}$, we can determine $\mathring{D}^i{}_j$ and $\mathring{D}$ as follows,
\allowdisplaybreaks
\begin{align}
    \mathring{D}^j{}_{i} &= 48 \bigg( \Box \xi_i{}^\mathcal{A}+ Y^a_i Y_a^k \xi_k{}^{\mathcal{A}} + \frac{1}{4} D_a (\bar{\Lambda}_R \gamma_a \Lambda_L \xi_i{}^{\mathcal{A}} ) + \frac{1}{4} \bar{\Lambda}_R \cancel{D} \xi_i{}^{\mathcal{A}} \Lambda_L + \frac{1}{4} F^{- \mathcal{A}} \cdot \mathring{T}_i - \frac{1}{16} \bar{\Lambda}_L \gamma \cdot \mathring{T}_i \theta_L^{\mathcal{A}} - \frac{1}{8} \xi^{j \mathcal{A}} \mathring{T}_j \cdot \mathring{T}_i  \nonumber \\
    &+ \frac{1}{24} \mathring{\bar{\zeta}}_i \theta_R ^\mathcal{A} + \frac{1}{16} E_i \bar{\Lambda}_R \theta_R^\mathcal{A} -\frac{1}{16} \mathring{\bar{\chi}}_{ik} \psi^{k \mathcal{A}} - \frac{1}{16} \mathring{\bar{\chi}}_i \psi_L^\mathcal{A}- \frac{1}{48} \varepsilon_{ik} \mathring{\bar{\zeta}}_L \psi^{k \mathcal{A}} + \frac{1}{48} \varepsilon_{ik} \mathring{\bar{\zeta}} ^k \psi_L^{ \mathcal{A}} \nonumber \bigg) \xi_{\mathcal{A}}{}^j+ \text{h.c}  - \text{trace} \;,\nonumber\\
    \mathring{D} &= 48 \bigg( \Box \xi_i{}^\mathcal{A}+ Y^a_i Y_a^j \xi_j^{\mathcal{A}} + \frac{1}{4} D_a (\bar{\Lambda}_R \gamma_a \Lambda_L \xi_i{}^{\mathcal{A}} ) + \frac{1}{4} \bar{\Lambda}_R \cancel{D} \xi_i{}^{\mathcal{A}} \Lambda_L + \frac{1}{4} F^{- \mathcal{A}} \cdot \mathring{T}_i - \frac{1}{16} \bar{\Lambda}_L \gamma \cdot \mathring{T}_i \theta_L^{\mathcal{A}} - \frac{1}{8} \xi^{i \mathcal{A}} \mathring{T}_i \cdot \mathring{T}_j \nonumber \\
    & + \frac{1}{24} \mathring{\bar{\zeta}}_i \theta_R ^\mathcal{A} + \frac{1}{16} E_i \bar{\Lambda}_R \theta_R^\mathcal{A} -\frac{1}{16} \mathring{\bar{\chi}}_{ij} \psi^{j \mathcal{A}}  - \frac{1}{16} \mathring{\bar{\chi}}_i \psi_L^\mathcal{A}- \frac{1}{48} \varepsilon_{ij} \mathring{\bar{\zeta}}_L \psi^{j \mathcal{A}} + \frac{1}{48} \varepsilon_{ij} \mathring{\bar{\zeta}} ^j \psi_L^{ \mathcal{A}}- \frac{1}{96} \xi_i{}^\mathcal{A} E^j E_j  \nonumber \\
    &- \frac{1}{96} \xi_i{}^\mathcal{A} \lvert E \rvert ^2+ \frac{1}{12} \xi_i{}^\mathcal{A} (\bar{\Lambda}_R \cancel{D}\Lambda_L + \bar{\Lambda}_L \cancel{D}\Lambda_R) + \frac{1}{12} \xi_i {}^\mathcal{A} \bar{\Lambda}_R \Lambda_R \bar{\Lambda}_L \bar{\Lambda}_L \bigg) \xi_{\mathcal{A}}{}^i + \text{h.c} \;.
\end{align}
The field equations of $\xi_{i}{}^{\AAA}$ have four complex components or eight real components. We have used four real components of these equations to solve for $\mathring{D}^{i}{}_{j}$ and $\mathring{D}$ respectively above. To analyze the remaining four components, we first contract the equation of motion of $\xi_i{}^{\mathcal{A}}$ with $\xi^{i{\mathcal{B}}} $. 

\allowdisplaybreaks
\begin{align} \label{eom xi}
    \xi^{i \mathcal{B}}\bigg( & \Box \xi_i{}^\mathcal{A}+ E^a_i E_a^j \xi_j{}^{\mathcal{A}} +\frac{1}{4} D_a (\bar{\Lambda}_R \gamma^a \Lambda_L \xi_i{}^{\mathcal{A}} )  + \frac{1}{4} \bar{\Lambda}_R \cancel{D}\xi_i{}^\mathcal{A}\Lambda_L + \frac{1}{4} F^{- \mathcal{A}} \cdot \mathring{T}_i - \frac{1}{16} \bar{\Lambda_L} \gamma \cdot \mathring{T}_i \theta_L^{\mathcal{A}} - \frac{1}{8} \xi^{j \mathcal{A}} \mathring{T}_i \cdot \mathring{T}_j  \nonumber \\
    &+ \frac{1}{24} \mathring{\bar{\zeta}}_i \theta_R ^\mathcal{A} + \frac{1}{16} E_i \bar{\Lambda}_R \theta_R^\mathcal{A}  -\frac{1}{16} \mathring{\bar{\chi}}_{ij} \psi^{j \mathcal{A}}  - \frac{1}{16} \mathring{\bar{\chi}}_i \psi_L^\mathcal{A} - \frac{1}{48} \varepsilon_{ij} \mathring{\bar{\zeta}}_L \psi^{j \mathcal{A}}+ \frac{1}{48} \varepsilon_{ij} \mathring{\zeta} ^j \psi_L^{ \mathcal{A}} - \frac{1}{48} \mathring{D}^j{}_{i} \xi_j^\mathcal{A} + \frac{1}{96} \mathring{D} \xi_i{}^{\mathcal{A}}\nonumber \\
 &- \frac{1}{96} \xi_i{}^\mathcal{A} \lvert E \rvert
 ^2 - \frac{1}{96} \xi_i{}^\mathcal{A} E^j E_j+ \frac{1}{12} \xi_i{}^\mathcal{A} (\bar{\Lambda}_R \cancel{D}\Lambda_L + \bar{\Lambda}_L \cancel{D}\Lambda_R) + \frac{1}{12} \xi_i {}^\mathcal{A} \bar{\Lambda}_R \Lambda_R \bar{\Lambda}_L \bar{\Lambda}_L\bigg)=0 \;.
\end{align}
\end{widetext}
\begin{widetext}
Consider the part of this equation that is symmetric in $\mathcal{A}$ and $\mathcal{B}$ and anti-hermitian, which have three real components,
\begin{align}
             \xi^{i( \mathcal{B} } \bigg(& \Box \xi_i{}^\mathcal{A)} +\frac{1}{4} D_a (\bar{\Lambda}_R \gamma^a \Lambda_L \xi_i{}^{\mathcal{A}} )  + \frac{1}{4} \bar{\Lambda}_R \cancel{D}\xi_i{}^\mathcal{A}\Lambda_L+ \frac{1}{4} F^{- \mathcal{A)}} \cdot \mathring{T}_i - \frac{1}{16} \bar{\Lambda_L} \gamma \cdot \mathring{T}_i \theta_L^{\mathcal{A)}} - \frac{1}{8} \xi^{i \mathcal{A)}}\mathring{T}_i \cdot \mathring{T}_j + \frac{1}{24} \bar{\zeta}_i \theta_R ^\mathcal{A)}  \nonumber \\
             & + \frac{1}{16} E_i \bar{\Lambda}_R \theta_R^\mathcal{A)} -\frac{1}{16} \bar{\chi}_{ij} \psi^{j \mathcal{A)}} - \frac{1}{16} \bar{\chi}_i \psi_L^\mathcal{A)} - \frac{1}{48} \varepsilon_{ij} \bar{\zeta}_L \psi^{j \mathcal{A)}} + \frac{1}{48} \varepsilon_{ij} \zeta ^j \psi_L^{ \mathcal{A)}} \bigg) - \text{h.c}  =0 \;.
\end{align}
This equation can be rewritten as 
\begin{align} \label{bianchi}
    &D_a \bigg( \xi^{i \mathcal{(B}} D^a \xi_i{}^{\mathcal{A})} - \xi_{i}^{(\mathcal{B}} D^a \xi^{i \mathcal{A)}} - \frac{1}{2} \bar{\theta}_L^{\mathcal{(A}} \gamma^a \theta_R^{\mathcal{B})} + \frac{1}{2} \bar{\psi}^{j \mathcal{(A}} \gamma^{a} \psi_j{}^{\mathcal{B})}  + \frac{1}{2} \bar{\psi}_L ^{(\mathcal{A}} \gamma^{a} \psi_R^{\mathcal{B})} +\frac{1}{2} \xi^{i(\mathcal{A}} \xi_{i}{}^{\mathcal{B})} \bar{\Lambda}_R \gamma^a \Lambda_L \bigg)\nonumber \\
   &= \frac{1}{8} F^{\mathcal{A}} \cdot \Tilde{F}^{\mathcal{B}} + \frac{1}{8} G ^{\mathcal{A}} \cdot \Tilde{G}^{\mathcal{B}} \;. 
\end{align}
Let us define the following.
\begin{align} \label{H def}
    \frac{1}{3!} \epsilon^{abcd} H_{bcd}^{\mathcal{A} \mathcal{B}}& \equiv  \xi^{i \mathcal{(B}} D^a \xi_i{}^{\mathcal{A})} - \xi_{i}{}^{(\mathcal{B}} D^a \xi^{i \mathcal{A)}} - \frac{1}{2} \bar{\theta}_L^{\mathcal{(A}} \gamma^a \theta_R^{\mathcal{B})} + \frac{1}{2} \bar{\psi}^{j \mathcal{(A}} \gamma^{a} \psi_j{}^{\mathcal{B})}+ \frac{1}{2} \bar{\psi}_L ^{(\mathcal{A}} \gamma^{a} \psi_R^{\mathcal{B})}  +\frac{1}{2} \xi^{i(\mathcal{A}} \xi_{i}{}^{\mathcal{B})} \bar{\Lambda}_R \gamma^a \Lambda_L \;.
\end{align}
With this definition \eqref{bianchi} becomes,
\begin{align}\label{Hequation}
    D_a \bigg( \frac{1}{3!} \epsilon^{abcd} H_{bcd}^{\mathcal{AB}}  \bigg)= \frac{1}{8} F^{\mathcal{A}} \cdot \Tilde{F}^{\mathcal{B}} + \frac{1}{8} G ^{\mathcal{A}} \cdot \Tilde{G}^{\mathcal{B}} \;.
\end{align}
\end{widetext}
By re-writing equation-\eqref{bianchi} as \eqref{Hequation}, we can interpret it as the Bianchi identity of a three-form field strength $H_{abc}^{\mathcal{A} \mathcal{B}}$ corresponding to a two-form gauge field $B_{\mu \nu}^{\mathcal{A} \mathcal{B}}$.

From the definition \eqref{H def} one can obtain the supersymmetry transformations of the field strength. They are given below.
\begin{widetext}
\begin{align} \label{Q transform of H}
    \delta_Q \bigg( \frac{1}{3!}  \epsilon_{abcd}H^{bcd, \mathcal{A} \mathcal{B}} \bigg) = & \bar{\epsilon}_i  \gamma_{ab} D^b(\theta_R^{(\mathcal{A}} \xi^{i \mathcal{B})}) +  \varepsilon_{ij} \bar{\epsilon}_L  \gamma_{ab} D^b(\bar{\psi}^{j \mathcal{(A}} \xi^{i \mathcal{B})}) -  \varepsilon_{ij} \bar{\epsilon}^j  \gamma_{ab} D^b(\bar{\psi}_L^{(\mathcal{A}} \xi^{i \mathcal{B})}) \nonumber \\
    +& \frac{1}{16} \varepsilon_{ij} \bar{\theta}^{(\mathcal{A}} \gamma \cdot \bigg(\mathring{T}^j \gamma_a \epsilon_L - T^+ \gamma_a \epsilon^j \bigg) \xi^{i \mathcal{B})} \nonumber\\
    +& \frac{1}{8} \bar{\psi}^{j ( \mathcal{A}} \gamma \cdot \mathring{T}_{[i} \gamma_a \epsilon_{j]} \xi^{i \mathcal{B})} 
    - \frac{1}{16} \bar{\psi}_L^{(\mathcal{A}} \gamma \cdot \bigg( \mathring{T}_i \gamma_a \epsilon_R - T^- \gamma_a \epsilon_i \bigg) \xi^{i \mathcal{B})} \nonumber \\
     -& \theta_R^{(\mathcal{A}} \gamma_{ab} \epsilon_R \xi^{i \mathcal{B})} E^b_i 
    + \varepsilon_{ij} \bar{\psi}^{j( \mathcal{A}} \gamma_{ab} \epsilon^k \xi^{i \mathcal{B})} E^b_k 
    +\varepsilon_{ij} \bar{\psi}_L ^{(\mathcal{A}} \gamma_{ab} \epsilon_L \xi^{i \mathcal{B})} E^{bj} \nonumber \\
    +& \frac{1}{8} \bar{\epsilon}^i  \gamma_a \gamma \cdot \bigg( F^{(\mathcal{A}} -iG^{(\mathcal{A}} \bigg) \psi_i^{\mathcal{B})} + \frac{1}{8} \bar{\epsilon}_L  \gamma_a  \gamma \cdot \bigg( F^{(\mathcal{A}} -i G^{(\mathcal{A}} \bigg)   \psi_{R}^{\mathcal{B})}  \nonumber \\
    +& \frac{1}{8} \bar{\epsilon}_i  \gamma_a  \gamma \cdot \bigg( F^{(\mathcal{A}}-iG^{(\mathcal{A}} \bigg) \Lambda_L \xi^{i \mathcal{B})}+ \frac{1}{8} \bar{\epsilon}^i   \gamma \cdot \bigg( F^{(\mathcal{A}} -iG^{(\mathcal{A}} \bigg) \gamma_a \psi_i^{\mathcal{B})} \nonumber\\
    +& \frac{1}{8} \bar{\epsilon}_L   \gamma \cdot \bigg( F^{(\mathcal{A}} -i G^{(\mathcal{A}} \bigg) \gamma_a  \psi_{R}^{\mathcal{B})} + \frac{1}{8} \bar{\epsilon}_i   \gamma \cdot \bigg( F^{(\mathcal{A}}-iG^{(\mathcal{A}} \bigg) \gamma_a \Lambda_L \xi^{i \mathcal{B})}  \nonumber\\ -&\text{h.c} \;, \nonumber \\
\delta_S \bigg( \frac{1}{3!}  \epsilon_{abcd}H^{bcd, \mathcal{A} \mathcal{B}} \bigg) &= -\frac{3}{2} \xi^{i \mathcal{(A}} \bar{\theta}_R^{\mathcal{B)}} \gamma_a \eta_i + \frac{3}{2} \varepsilon_{ij} \xi^{i \mathcal{(A}} \bar{\psi}_L^{\mathcal{B)}} \gamma_a \eta^j - \frac{3}{2} \varepsilon_{ij} \xi^{i \mathcal{(A}} \bar{\psi}^{j \mathcal{B})} \gamma_a \eta_R \;.
\end{align} 
\end{widetext}
From the bianchi identity \eqref{bianchi}, supersymmetry transformations \eqref{Q transform of H} and the knowledge of the superconformal soft algebra, we can obtain the gauge transformations of the two-form gauge field $B_{\mu \nu}^{\mathcal{A} \mathcal{B}} $, as given in \eqref{B}. This gauge field comes with its own  vector gauge transformation parametrized by $\Lambda_\mu^{\mathcal{AB}}$. As we can see in equation-\eqref{B}, the 2-form gauge field $B_{\mu \nu}^{\mathcal{A}\mathcal{B}}$ also transforms non-trivially under the $U(1)$ gauge transformations associated with the gauge fields $C_{\mu}^{\AAA}$ and $\Tilde{C}_{\mu}^{\AAA}$ parametrized by $\lambda^{\AAA}$ and $\Tilde{\lambda}^{\AAA}$ respectively.
\begin{widetext}
\begin{align} \label{B}
    \delta B_{\mu \nu}^{\mathcal{A} \mathcal{B} }=& 2 \partial_{[\mu} \Lambda_{\nu]}^{\mathcal{AB}}+ \frac{1}{2} C_{[\mu}^{(\mathcal{A}} \delta C_{\nu]}^{\mathcal{B})} + \frac{1}{2} \Tilde{C}^{(\mathcal{A}}_{[\mu} \delta \Tilde{C}_{\nu]}^{\mathcal{B})} - \frac{1}{4} \lambda^{(\mathcal{A}} F_{\mu \nu}^{\mathcal{B})} - \frac{1}{4} \Tilde{\lambda}^{(\mathcal{A}} G_{\mu \nu}^{\mathcal{B})}+\Bigg( \bar{\epsilon}_i \gamma_{\mu \nu} \theta_R^{(\mathcal{A}} \xi^{i \mathcal{B})}-\varepsilon_{ij} \bar{\epsilon}_L \gamma_{\mu \nu} \psi^{j (\mathcal{A}} \xi^{i \mathcal{B})}\nonumber \\
    +& \varepsilon_{ij} \bar{\epsilon}^j \gamma_{\mu \nu} \psi_L^{(\mathcal{A}} \xi^{i \mathcal{B})}
    -4 \xi_i^{(\mathcal{A}} \xi^{j \mathcal{B)}} \bar{\epsilon}^i \gamma_{[\mu} \psi_{\nu] j}+ 2 \xi^{i \mathcal{(A}} \xi_i{}^{\mathcal{B)}} \bar{\epsilon}_L \gamma_{[\mu} \psi_{\nu]R} +2\xi^{i \mathcal{(A}} \xi_i{}^{\mathcal{B)}} \bar{\epsilon}^j \gamma_{[\mu} \psi_{\nu] j} +\text{h.c}\Bigg) \;.
\end{align}
From the above transformations, the explicit form of the 3-form field strength $H_{\mu\nu\rho}^{\AAA\BB}$ can be obtained as follows:
\begin{align}
    H_{\mu\nu\rho}^{\mathcal{A} \mathcal{B}}&= 3 \partial_{[\mu} B_{\nu\rho]}^{\mathcal{A} \mathcal{B}} +\frac{3}{4} C^{\mathcal{(A}}_{[\mu}F_{\nu\rho]}^{\mathcal{B})} + \frac{3}{4} \Tilde{C}_{[\mu}^{(\mathcal{A}} G_{\nu\rho]}^{\mathcal{B})}-\Bigg(\frac{3}{2} \bar{\psi}_{[\mu i} \gamma_{\nu\rho]} \theta_R^{(\mathcal{A}} \xi^{i \mathcal{B})} +\frac{3}{2} \varepsilon_{ij} \bar{\psi}_{[\mu L} \gamma_{\nu\rho]} \psi^{j (\mathcal{A}} \xi^{i \mathcal{B})} -  \frac{3}{2} \varepsilon_{ij} \bar{\psi}_{[\mu}^j \gamma_{\nu\rho]} \psi_L^{\mathcal{(A}} \xi^{i \mathcal{B)}} \nonumber \\
    & +6 \xi_i^{\mathcal{(A}} \xi^{j \mathcal{B})} \psi^i_{[\mu} \gamma_{\nu} \psi_{\rho]j} -3 \xi^{i \mathcal{(A}} \xi_i{}^{\mathcal{B})} \psi_{[\mu L} \gamma_{\nu} \psi_{\rho]R} - 3 \xi^{i \mathcal{(A}}\xi_i{}^{\mathcal{B})} \bar{\psi}_{[\mu}^j \gamma_{\nu} \psi_{\rho]j} +\text{h.c}\Bigg) \;.
\end{align}
\end{widetext}
Now, consider the part of \eqref{eom xi} that is antisymmetric in $\mathcal{A}$ and $\mathcal{B}$ and real, which contains one real component. By contracting this component with the Levi-Civita $\epsilon_{\AAA\BB}$, we can re-write it as,
\begin{widetext}
\begin{align} \label{bianchi M}
   \epsilon_{\mathcal{A} \mathcal{B}}  D_{a} \bigg( \xi^{i \mathcal{B}} D^a \xi_{i}{}^{\mathcal{A}} + \xi_i{}^{\mathcal{B}} D^a \xi^{i \mathcal{A}} + \frac{1}{2} \bar{\theta}_R^{\mathcal{A}} \gamma^a \theta_L^{\mathcal{B}} + \frac{1}{2} \bar{\psi}^{j \mathcal{A}} \gamma^a \psi_j^{\mathcal{B}}  + \frac{1}{2} \bar{\psi}^{\mathcal{A}}_L \gamma^a \psi_R^{\mathcal{B}} + \frac{1}{2} \xi^{i \mathcal{B}} \xi_i{}^{\mathcal{A}} \bar{\Lambda}_R \gamma^a \Lambda_L \bigg)= \frac{i}{4} \epsilon_{\mathcal{A} \mathcal{B}} F^{\mathcal{A}} \cdot \Tilde{G}^{B} 
\end{align}
\end{widetext}
Let us define a 3-form $M_{abc}$ as follows:
\begin{align} \label{M}
     \frac{i}{3!} \varepsilon^{abcd}M_{bcd}\nonumber \equiv\epsilon_{\mathcal{A} \mathcal{B}}  \bigg(& \xi^{i \mathcal{B}} D^a \xi_{i}{}^{\mathcal{A}} + \xi_i{}^{\mathcal{B}} D^a \xi^{i \mathcal{A}} + \frac{1}{2} \bar{\theta}_R^{\mathcal{A}} \gamma^a \theta_L^{\mathcal{B}} \nonumber \\
     &+ \frac{1}{2} \bar{\psi}^{j \mathcal{A}} \gamma^a \psi_j^{\mathcal{B}}+ \frac{1}{2} \bar{\psi}^{\mathcal{A}}_L \gamma^a \psi_R^{\mathcal{B}} \nonumber \\
     &+\frac{1}{2} \xi^{i \mathcal{B}} \xi_i{}^{\mathcal{A}} \bar{\Lambda}_R \gamma^a \Lambda_L \bigg) \;.
\end{align}
With this definition, \eqref{bianchi M} can be re-written as,
\begin{align} \label{realbianchiM}
    D_a \bigg(\frac{i}{3!} \varepsilon^{abcd}M_{bcd}  \bigg)= \frac{i}{4} \epsilon_{\mathcal{AB}} F^{\mathcal{A}} \cdot \tilde{G}^{\mathcal{B}} \;.
\end{align}
We can interpret the above equation \eqref{realbianchiM} as the Bianchi identity of the three-form field strength $M^{abc}$ corresponding to a new two-form gauge field denoted as $E_{\mu \nu}$
\begin{widetext}
  From the definition \eqref{M}, we can obtain the supersymmetry transformations of the field strength $M^{abc}$ which is given below,
\begin{align} \label{trasnform M}
    \delta_Q \bigg( \frac{i}{3!} \epsilon^{abcd} M_{bcd} \bigg)=&-\epsilon_{\mathcal{A} \mathcal{B}} \bar{\epsilon}_i \gamma^{ab} D_b ( \theta_R^{\mathcal{B}} \xi^{i \mathcal{A}}) - \epsilon_{\mathcal{A} \mathcal{B}} \varepsilon_{ij} \bar{\epsilon}_L \gamma^{ab} D_b( \psi^j \xi^{\mathcal{A}} ) + \epsilon_{\mathcal{A} \mathcal{B}} \varepsilon_{ij} \bar{\epsilon}^j \gamma^{ab} D_b (\psi_L^{\mathcal{B}}  \xi^{i \mathcal{A}}) \nonumber\\
    &- \frac{1}{16} \epsilon_{\mathcal{A} \mathcal{B}} \varepsilon_{ij} \xi^{i \mathcal{A}} \bar{\theta}_R^{\mathcal{B}}  \gamma \cdot \bigg(\mathring{T}^j \gamma^a \epsilon_L - T^+ \gamma^a \epsilon^j \bigg) - \frac{1}{8} \epsilon_{\mathcal{A} \mathcal{B}} \xi^{i \mathcal{A}} \bar{\psi}^{j \mathcal{B}}  \gamma \cdot \mathring{T}_{[i} \gamma^a \epsilon_{j]}  \nonumber \\
    & -\frac{1}{16} \epsilon_{\mathcal{A} \mathcal{B}} \xi^{i \mathcal{A}} \bar{\psi}^{\mathcal{B}}_L \gamma \cdot \bigg( \mathring{T}_i \gamma^a \epsilon_R - T^- \gamma^a \epsilon_i \bigg) + \epsilon_{\mathcal{A} \mathcal{B}} \xi^{i \mathcal{A}} \bar{\theta}_R^{\mathcal{B}} \gamma^{ab} \epsilon_R E_{bi} \nonumber \\
    &+ \epsilon_{\mathcal{A} \mathcal{B}} \varepsilon_{ij} \xi^{i \mathcal{A}} \bar{\psi}^{j \mathcal{B}} \gamma^{ab} \epsilon^k E_{bk} - \epsilon_{\mathcal{A} \mathcal{B}} \varepsilon_{ij} \xi^{i \mathcal{A}} \bar{\psi}_L^{\mathcal{B}} \gamma^{ab} E_b^j \epsilon_L \nonumber \\
&{ - \frac{1}{8} \epsilon_{\mathcal{A} \mathcal{B}} \bar{\psi}_L^{\mathcal{A}} \gamma^a \gamma \cdot \bigg( F^{ \mathcal{B}+} + i G^{\mathcal{B}+} \bigg) \epsilon_R - \frac{1}{8} \epsilon_{\mathcal{A} \mathcal{B}} \bar{\psi}_L^{\mathcal{A}} \gamma \cdot \bigg( F^{\mathcal{B}-} + iG^{\mathcal{B} -}  \bigg) \epsilon_R } \nonumber\\
&{ - \frac{1}{8} \epsilon_{\mathcal{A} \mathcal{B}} \bar{\psi}^{i\mathcal{A}} \gamma^a \gamma \cdot \bigg( F^{ \mathcal{B}+} + i G^{\mathcal{B}+} \bigg) \epsilon_i - \frac{1}{8} \epsilon_{\mathcal{A} \mathcal{B}} \bar{\psi}^{i \mathcal{A}} \gamma \cdot \bigg( F^{\mathcal{B}-} + iG^{\mathcal{B} -}  \bigg) \epsilon_i } \nonumber \\
& { - \frac{1}{8} \epsilon_{\mathcal{A} \mathcal{B}} \xi^{i \mathcal{A}}  \bar{\Lambda}_L\gamma^a \gamma \cdot \bigg( F^{ \mathcal{B}+} - i G^{\mathcal{B}+} \bigg) \epsilon_i - \frac{1}{8} \epsilon_{\mathcal{A} \mathcal{B}} \xi^{i \mathcal{A}} \bar{\Lambda}_L \gamma \cdot \bigg( F^{\mathcal{B}-} - iG^{\mathcal{B} -}  \bigg) \epsilon_i } \nonumber \\
&+\text{h.c} \;, \nonumber \\
\delta_S \bigg( \frac{i}{3!} \epsilon^{abcd}M_{bcd}  \bigg)=& -\frac{3}{2} \epsilon_{\mathcal{AB}} \xi^{i \mathcal{B}} \bar{\theta}_R^{\mathcal{A}} \gamma^a \eta_i + \frac{3}{2} \epsilon_{\mathcal{AB}} \varepsilon_{ij} \xi^{i \mathcal{B}} \bar{\psi}_L^{\mathcal{A}} \gamma^a \eta^j -\frac{3}{2} \epsilon_{\mathcal{AB}} \varepsilon_{ij} \xi^{i \mathcal{B}} \bar{\psi}^{j \mathcal{A}} \gamma^a \eta_L + \text{h.c} \;.
\end{align}  
\end{widetext}

From the Bianchi identity \eqref{realbianchiM}, supersymmetry transformations of the field strength \eqref{trasnform M}, and the superconformal algebra, one can find the gauge transformations of the gauge field $E_{\mu \nu}$. The result is given in \eqref{E}. The 2-form $E_{\mu \nu}$ comes with its own vector gauge transformation parametrized by $\sigma_{\nu}$. It also has non trivial $U(1)$ gauge transformations corresponding to the gauge fields $C_{\mu}$ and $\Tilde{C}_{\mu}$. 
\begin{widetext}
\begin{align} \label{E}
    \delta E_{\mu \nu}=2 \partial_{[\mu} \sigma_{\nu]}& + \frac{1}{2} \epsilon_{\mathcal{A} \mathcal{B}} \Tilde{C}^{\mathcal{B}}_{[\mu} \delta C^{\mathcal{A}}_{\nu]} + \frac{1}{2} \epsilon_{\mathcal{A} \mathcal{B}} C^{\mathcal{A}}_{[\mu} \delta \Tilde{C}_{\nu ]}^{\mathcal{B}} - \frac{1}{4} \epsilon_{\mathcal{A} \mathcal{B}} \Tilde{\lambda}^{\mathcal{B}} F^{\mathcal{A}}_{\mu \nu} -\frac{1}{4} \epsilon_{\mathcal{A} \mathcal{B}} \lambda^{\mathcal{A}} G^{\mathcal{B}}_{\mu \nu} \nonumber \\
    +\epsilon_{\mathcal{A} \mathcal{B}}\bigg(&i \bar{\epsilon}_i \gamma_{\mu \nu} \theta_R^{\mathcal{B}} \xi^{i \mathcal{A}}  -i \varepsilon_{ij} \bar{\epsilon}_L \gamma_{\mu \nu} \psi^{j \mathcal{B}} \xi^{i \mathcal{A}}+ i \varepsilon_{ij} \bar{\epsilon}^j \gamma _{\mu \nu} \psi_L^{\mathcal{B}} \xi^{i \mathcal{A}}-4i \xi^{i\mathcal{A}} \xi_{j}{}^{\mathcal{B}} \bar{\epsilon}_i \gamma_{[\mu} \psi_{\nu]}^j \nonumber \\
    &  + 2i \xi^{i\mathcal{A}} \xi_i^{\mathcal{B}} \bar{\epsilon}^j \gamma_{[\mu} \psi_{\nu]j} + 2i \xi^{i \mathcal{A}} \xi_i{}^{\mathcal{B}} \bar{\epsilon}_L \gamma_{[\mu} \psi_{\nu] L} +\text{h.c}\bigg) \;.
\end{align}
From the above gauge transformations of the 2-form $E_{\mu\nu}$, the explicit form of the field strength $M_{\mu\nu\rho}$ can be obtained. It is as follows,
\begin{align}
    M_{\mu\nu\rho}&= 3 \partial_{[\mu} E_{\nu\rho]} + \frac{3}{4} \epsilon_{\mathcal{A} \mathcal{B}} \Tilde{C}_{[\mu}{}^{\mathcal{B}} F_{\nu\rho]}{}^{\mathcal{A}} + \frac{3}{4} \epsilon_{\mathcal{A} \mathcal{B}} C_{[\mu} {}^{\mathcal{A}}G_{\nu\rho]}{}^{\mathcal{B}} -\frac{3i}{2} \epsilon_{\mathcal{AB}} \bar{\psi}_{[\mu i} \gamma_{\nu\rho]}\theta_R^{\mathcal{B}} \xi^{i\mathcal{A}}+ \frac{3i}{2} \epsilon_{\mathcal{AB}}\varepsilon_{ij} \bar{\psi}_{[\mu L} \gamma_{\nu\rho]} \psi^{j \mathcal{B}} \xi^{i \mathcal{A}} \nonumber \\
    &+ \epsilon_{\mathcal{AB}} \bigg( +6i \xi^{\mathcal{A}} \xi_j{}^{\mathcal{B}} \bar{\psi}_{[\mu i} \gamma_{\nu} \psi_{\rho]}^j - 3i \xi^{i \mathcal{A}} \xi_i{}^{\mathcal{B}} \bar{\psi}_{[\mu}^j \gamma_{\nu} \psi_{\rho]j} -3i\xi^{i \mathcal{A}} \xi_i{}^{\mathcal{B}} \bar{\psi}_{[\mu L} \gamma_{\nu} \psi_{\rho]L}+ \text{h.c} \bigg)
\end{align}
\end{widetext}
We have solved all the remaining four components of the scalar field equations by re-writing them as the Bianchi identities of various 3-form field strengths. In the next few steps, we would like to show that by doing so, we have actually solved for the auxiliary gauge fields $\mathring{v}_{\mu}{}^{i}{}_{j}$ and $\mathring{v}_{\mu}$ corresponding to the R-symmetry $SU(2)\times U(1)_v$ in terms of these 3-form field strengths.
\begin{widetext}
By extracting the $SU(2)\times U(1)_v$ gauge fields from the R.H.S of the definition of the 3-form field strength $H_{abc}^{\AAA\BB}$ \eqref{H def}, we can get the following:
\begin{align} \label{V first step}
    2 \mathring{v}_a{}^{j}{}_{i} \xi^{i (\mathcal{A}} \xi_j{}^{\mathcal{B})} - i \mathring{v}_a \xi^{i (\mathcal{A}} \xi_i{} ^{\mathcal{B})} &=  \frac{1}{3!} \epsilon_{abcd}H^{bcd \mathcal{A} \mathcal{B}}-\bigg( \xi^{i \mathcal{(B}} D'^a \xi_i{}^{\mathcal{A})} - \xi_{i}{}^{(\mathcal{B}} D'^{a} \xi^{i \mathcal{A)}} - \frac{1}{2} \bar{\theta}_L^{\mathcal{(A}} \gamma^a \theta_R^{\mathcal{B})} \nonumber \\ &+ \frac{1}{2} \bar{\psi}^{j \mathcal{(A}} \gamma^{a} \psi_j{}^{\mathcal{B})} + \frac{1}{2} \bar{\psi}_L ^{(\mathcal{A}} \gamma^{a} \psi_R^{\mathcal{B})} + \frac{1}{2} \xi^{i(\mathcal{A}} \xi_{i}{}^{\mathcal{B})} \bar{\Lambda}_R \gamma^a \Lambda_L \bigg)
\end{align}
\end{widetext}
Where the derivative $D'_a$ is covariant with respect to all the gauge symmetries except the R-symmetry $SU(2)\times U(1)_v$. Similarly by expanding the covariant derivatives in the R.H.S of the definition of the 3-form field strength $M_{abc}$ \eqref{M} we can get the following:
\begin{widetext}
\begin{align} \label{V step 2}
      2 \mathring{v}{}_a{}^{j}{}_{i} \xi^{i [\mathcal{B} }\xi_j{}^{\mathcal{A}]} - i \mathring{v}_a \xi^{i \mathcal{[B}} \xi_i{}^{\mathcal{A}]} &=
       \frac{i}{12} \epsilon^{\mathcal{A} \mathcal{B}} \varepsilon^{abcd}M_{bcd}  - \bigg(\xi^{i [ \mathcal{B}}  D'^a \xi_{i}{}^{\mathcal{A}]} + \xi_i{}^{[\mathcal{B}} D'^a \xi^{i \mathcal{A}]} + \frac{1}{2} \bar{\theta}_R^{[\mathcal{A}} \gamma^a \theta_L^{\mathcal{B}]}\nonumber
       \\& + \frac{1}{2} \bar{\psi}^{j \mathcal{[A}} \gamma^a \psi_j^{\mathcal{B}]}+ \frac{1}{2} \bar{\psi}^{[\mathcal{A}}_L \gamma^a \psi_R^{\mathcal{B}]} + \frac{1}{2} \xi^{i[ \mathcal{B}} \xi_i{}^{\mathcal{A}]} \bar{\Lambda}_R \gamma^a \Lambda_L \bigg) 
\end{align}
\end{widetext}



By taking appropriate linear combinations of the equations-\eqref{V first step} and \eqref{V step 2}, one can obtain composite expressions for the $SU(2)$ gauge field $\mathring{v}_a{}^i{}_j$ and the $U(1)_v$ gauge field $\mathring{v}_a$ as shown below. 

\begin{align}
    \mathring{v}_a{}^{i}{}_j &= \frac{1}{2} J_a^{\mathcal{A} \mathcal{B}} (\xi_{\mathcal{A}j} \xi_{\mathcal{B}}{}^i - \frac{1}{2} \delta^i_j \xi_{\mathcal{A} k} \xi_{\mathcal{B}}{}^k )\;, \nonumber \\
\mathring{v}_a &= \frac{i}{2} J_a^{\mathcal{A} \mathcal{B}} \xi_{\mathcal{A}i} \xi_{\mathcal{B}}{}^i\;,
\end{align}
where we have defined:
\begin{widetext}
    \begin{align}
   J_{a}^{\AAA\BB}&= \frac{1}{6} \epsilon_{abcd}H^{bcd \mathcal{A} \mathcal{B}}
    + \frac{i}{12} \epsilon^{\mathcal{A} \mathcal{B}} \varepsilon^{abcd}M_{bcd} \nonumber \\
    &- \bigg( \xi^{i  \mathcal{B}}  D'^a \xi_{i}{}^{\mathcal{A}} + \xi_i{}^{\mathcal{B}} D'^a \xi^{i \mathcal{A}} + \frac{1}{2} \bar{\theta}_R^{\mathcal{A}} \gamma^a \theta_L^{\mathcal{B}} + \frac{1}{2} \bar{\psi}^{j \mathcal{A}} \gamma^a \psi_j{}^{\mathcal{B}}+ \frac{1}{2} \bar{\psi}^{\mathcal{A}}_L \gamma^a \psi_R^{\mathcal{B}} + \frac{1}{2} \xi^{i \mathcal{B}} \xi_i{}^{\mathcal{A}} \bar{\Lambda}_R \gamma^a \Lambda_L \bigg) 
    \end{align}
\end{widetext}
Now, we have exhausted all the field equations of the vector multiplets. The remaining independent fields of the standard Weyl multiplet, the vector multiplet fields and the newly introduced gauge fields in \eqref{tildeC}, \eqref{B} and \eqref{E} constitute a new off-shell multiplet, which is our variant $\mathcal{N}=3$ dilaton Weyl multiplet. In table \ref{3dilaton4d} we give the details of the field content of this dilaton Weyl multiplet.  
\begin{widetext}
 \begin{center}
\begin{table}
		\centering
		\begin{tabular}{|P{2cm}|P{4cm}|P{2cm}|P{2cm}|P{2cm}|}
			\hline
			Field&Properties&SU(2) irreps& $w$ & ($c_A, c_v$)\\
			\hline
			\multicolumn{5}{|c|}{Independent Gauge fields}\\
			\hline
			$e_\mu^a$&Vielbein&\bf{1}&$-1$&$(0,0)$\\
			$\psi_\mu^i$& $ \gamma_5 \psi_\mu^i=\psi_\mu^i$, gravitino  &\bf{2}&$-1/2$&$(-1/2,-1/2 )$\\
            $\psi_{\mu L}$& $\gamma_5 \psi_{\mu L}= \psi_{\mu L}$, gravitino & \bf{1}&$-1/2$&$(-1/2,1)$\\
             $b_\mu$&Dilatation gauge field&\bf{1}&$0$&$(0,0)$\\
   			$A_\mu$&$ U(1) $ gauge field&\bf{1}&0&$(0,0)$\\
            $C_\mu{}^\mathcal{A} $ & $ U(1) $ gauge field&\bf{1}&0&$(0,0)$\\
              $\Tilde{C}_\mu{}^\mathcal{A} $ & $ U(1) $ gauge field&\bf{1}&0&$(0,0)$\\
                $B_{\mu\nu}{}^\mathcal{A B} $ & Two-form gauge field&\bf{1}&0&$(0,0)$\\
                $E_{\mu\nu} $ & Two-form gauge field&\bf{1}&0&$(0,0)$\\
			\hline
			\multicolumn{5}{|c|}{Covariant matter fields}\\
			\hline
   $Y_a^i$&Boson&\bf{2}&1&$(0,-1/2)$\\
 $  T_{ab}^+$& $T^+_{ab}=\frac{1}{2}\varepsilon_{abcd}T^{+cd} $&\bf{1}&1&$(1,1)$\\
 $E_i$& Complex & $\bf{\Bar{2}}$ &1&$(-1,1/2)$\\
  $E$& Complex & $\bf{1}$ &1&$(-1,-1)$\\
$\chi_L$&$\gamma_5\chi_L=\chi_L $& $\bf{\Bar{1}}$&$3/2$&$(-1/2, -1)$\\
$\Lambda_L$&$\gamma_5\Lambda_L=\Lambda_L$&\bf{1}&$1/2$&$(-3/2,0)$\\
$\xi_i{}^\mathcal{A} $& Complex scalar & \bf{2} & $1$& $(-1, 1/2)$\\
$\psi_i{}^\mathcal{A}$&$\gamma_5 \psi_i^{\mathcal{A}}=-\psi_i^{\mathcal{A}}$ &\bf{2}&$3/2$&$(1/2,1/2)$\\
$\psi_R^\mathcal{A}$& $\gamma_5 \psi_R^{\mathcal{A}}=-\psi_R^{\mathcal{A}}$& \bf{1}&$3/2$&$(1/2,-1)$\\
$\theta_L{}^\mathcal{A}$& $\gamma_5 \theta_L^{\mathcal{A}}=\theta_L^{\mathcal{A}}$ & \bf{1}&$3/2$&$(3/2,0)$ \\
 			\hline
		\end{tabular}
		\caption{Field content of the variant $\mathcal{N}=3$ dilaton Weyl multiplet}
		\label{3dilaton4d}	
	\end{table}
    \end{center}   
The supersymmetry transformations of the component fields of this variant dilaton Weyl multiplet are given in equations-\eqref{dilaontrans}.
\begin{subequations}\label{dilaontrans}
{\begin{align}
\delta e_\mu^a&=\bar{\epsilon}_i\gamma^a\psi_\mu^i+\bar{\epsilon}_R\gamma^a\psi_{\mu L}+\text{h.c.} \\[5pt]
\delta\psi_\mu^i=&2\mathcal{D}_\mu\epsilon^i-2Y^i_\mu\epsilon_L-u(\psi_\mu)^i\epsilon_L-\frac{1}{8} \varepsilon^{ij} \gamma\cdot \bigg(\mathring{T}_j\gamma_\mu\epsilon_R- T^-\gamma_\mu\epsilon_j \bigg)\nonumber\\
    &-\varepsilon^{ij}\bar{\epsilon}_j\psi_{\mu R}\Lambda_L +\varepsilon^{ij}\bar{\epsilon}_R \psi_{\mu j}\Lambda_L+ u(\epsilon)^i\psi_{\mu L}-\gamma_\mu \eta^i \\[5pt]
     \delta \psi_{\mu L}=&2\mathcal{D}_\mu\epsilon_L+2Y_{\mu j}\epsilon^j + u(\psi_\mu)_i \epsilon^i -\frac{1}{8}\varepsilon^{ij}\gamma\cdot \mathring{T}_i\gamma_\mu \epsilon_j-\varepsilon^{ij}\bar{\epsilon}_i\psi_{\mu j}\Lambda_L-u(\epsilon)_i\psi^i_\mu 
   -\gamma_\mu\eta_R   \\
    \delta b_\mu=&\frac{1}{2}\bigg(\bar{\epsilon}^i\phi_{\mu i}+\bar{\epsilon}_L\phi_{\mu L}-\bar{\psi}_\mu^i\eta_i-\bar{\psi}_{\mu L}\eta_L \bigg)+\text{h.c.} \\[5pt]
    \delta A_\mu=& \frac{i}{6}\bigg( \bar{\epsilon}^i\phi_{\mu i}+\bar{\epsilon}_L\phi_{\mu L}\bigg)+\frac{i}{36}\bigg( \bar{\epsilon}^i\gamma_\mu \mathring{\zeta}_i+\bar{\epsilon}_L\gamma_\mu \mathring{\zeta}_R\bigg)+\frac{i}{12}\varepsilon_{kl}\bigg(\bar{E}\bar{\epsilon}^k\psi^l_\mu-E^l\bar{\epsilon}^k\psi_{\mu L}\nonumber\\
    &+E^l\bar{\epsilon}_L\psi^k_\mu
    \bigg) +\frac{i}{12}\bigg(\bar{\epsilon}^i\gamma\cdot \mathring{T}_i\gamma_\mu\Lambda_R+\bar{\epsilon}_L\gamma\cdot T^-\gamma_\mu \Lambda_R \bigg)+\frac{i}{12}\bigg(E_i\bar{\epsilon}^i\gamma_\mu\Lambda_R+E\bar{\epsilon}_L\gamma_\mu \Lambda_R \bigg) \nonumber\\
    &-\frac{i}{3}\bigg( \bar{\epsilon}^i\gamma^a\psi_{\mu i}+\bar{\epsilon}_L\gamma_a\psi_{\mu R}\bigg)\bar{\Lambda}_L\gamma_a\Lambda_R-\frac{i}{16}\bigg(\bar{\psi}_\mu^i\eta_i+\psi_{\mu L}\eta_L \bigg)+\text{h.c.}\\[5pt]
    \delta\Lambda_L=&-\frac{1}{4}\bigg(E_i\epsilon^i+E\epsilon_L \bigg)+\frac{1}{4}\gamma\cdot \bigg(\mathring{T}_i\epsilon^i+T^-\epsilon_L \bigg)\\[5pt]
    \delta E_i=&-4\bar{\epsilon}_i\cancel{D}\Lambda_L-\frac{1}{2}\varepsilon_{ij}\bigg(\bar{\epsilon}^j \mathring{\zeta}^L-\bar{\epsilon}_L \mathring{\zeta}^j \bigg)+\frac{1}{2}\bar{\epsilon}^j \mathring{\chi}_{ij}+\frac{1}{2}\bar{\epsilon}_L \mathring{\chi}_i-\frac{1}{2}\varepsilon_{ij}\bigg(\bar{E}\bar{\epsilon}^j\Lambda_L-E^j\bar{\epsilon}_L\Lambda_L \bigg) \nonumber\\
    &-4\bar{\Lambda}_L\Lambda_L\bar{\epsilon}_i\Lambda_R-4\bar{\eta}_i\Lambda_L+E u(\epsilon)_i\\
    \delta E=&-4\bar{\epsilon}_R\cancel{D}\Lambda_L-\frac{1}{2}\varepsilon_{jk}\bar{\epsilon}^j \mathring{\zeta}^k+\frac{1}{2}\bar{\epsilon}^j \mathring{\chi}_j+\frac{1}{2}\bar{\epsilon}_L \chi_L-\frac{1}{2}\varepsilon_{jk}E^k\bar{\epsilon}^j\Lambda_L-4\bar{\Lambda}_L\Lambda_L\bar{\epsilon}_R\Lambda_R\nonumber\\
    &-E_ju(\epsilon)^j-4\bar{\eta}_L\Lambda_L \\[5pt]
   \delta T^+_{ab} =& -\bar{\epsilon}_L \cancel{D}\gamma_{ab}\Lambda_R-4\varepsilon^{jk}\bar{\epsilon}_j R_{ab}(Q)_k+\frac{1}{8}\bar{\epsilon}_j\gamma_{ab} \mathring{\chi}^{j}+\frac{1}{8}\bar{\epsilon}_R \chi_R +\frac{1}{24}\varepsilon^{jk}\bar{\epsilon}_j\gamma_{ab} \mathring{\zeta}_k \nonumber \\
	&-\frac{1}{8}\varepsilon^{jk}E_j\bar{\epsilon}_k\gamma_{ab}\Lambda_R-u(\epsilon)_i \mathring{T}^i_{ab}+\bar{\eta}_R\gamma_{ab}\Lambda_R \\[5pt]
      \delta \chi_L =& 2 \cancel{D}E \epsilon_R + 2 \slashed{Y}^iE_i\epsilon_R -16 \varepsilon_{ij} \gamma^{ab} D_{[a}Y_{b]}^j \epsilon^i+\frac{1}{4} \varepsilon_{ij} \xi_{\mathcal{A}}{}^j \gamma^{ab} \bar{\theta}_R^{\mathcal{A}} R(Q)_{ab R}-\frac{1}{4}\varepsilon_{ij} \varepsilon_{kl} \xi_{\mathcal{A}}{}^j \gamma^{ab} \bar{\psi}^{l \mathcal{A}} R(Q)_{ab}^k\nonumber\\
      &-2\gamma \cdot \cancel{D}T^- \epsilon_R -2\gamma^{ab} \slashed{Y}^i \mathring{T}_{ab i} + \frac{1}{3} \varepsilon_{ij} \mathring{D}^j \epsilon^i + \frac{1}{4} \varepsilon_{ij} E^i \gamma \cdot T^- \epsilon^j - \frac{1}{3} \bar{\Lambda}_L \gamma_a \epsilon_R \gamma^a \mathring{\zeta}_R \nonumber\\
      &+\frac{1}{4} \varepsilon_{ij}EE^j \epsilon^i- \bar{\Lambda}_L \gamma^a \Lambda_R \gamma_a E \epsilon_R -\bar{\Lambda}_L \gamma \cdot T^- \gamma^a \Lambda_R \gamma_a \epsilon_R-2u(\epsilon)^i \mathring{\chi}_i 
     +2 \gamma \cdot T^- \eta_L + 2 E \eta_L
      \\[5pt]
    \delta Y_a^i =& -\bar{\epsilon}_R D_a(\theta_R^{\mathcal{A}} \xi_{\mathcal{A}}{}^i )+ \varepsilon_{jk} \bar{\epsilon}^j D_a(\psi^{k \mathcal{A}} \xi_{\mathcal{A}}{}^i ) - \frac{1}{48} \bar{\epsilon}^i \gamma_a \mathring{\zeta}_R + \frac{1}{48} \bar{\epsilon}_R \gamma_a \mathring{\zeta}^i + \frac{1}{16} \varepsilon_{jk} \bar{\epsilon}^j \gamma_a \mathring{\chi}^{ik} \nonumber\\
    & -\frac{1}{16}\varepsilon^{ij} \bar{\epsilon}_j \gamma_a \chi_L+ \frac{1}{16} \varepsilon^{ik} \bar{\epsilon}_R \gamma_a \mathring{\chi}_k- \frac{1}{16} \bar{\epsilon}^i \gamma \cdot T^- \gamma_a \Lambda_R + \frac{1}{16} \bar{\epsilon}_R \gamma \cdot T^i \gamma_a \Lambda_L -\frac{1}{16} \bar{\epsilon}^i \gamma_a \Lambda_R E \nonumber\\
    &+ \frac{1}{16} \bar{\epsilon}_R \gamma_a \Lambda_L E^i+\xi_{\mathcal{A}}{}^i Y_a^j \bar{\theta}_R^{\mathcal{A}} \epsilon_j -\frac{1}{16} \varepsilon_{jk}\xi_{\mathcal{A}}{}^i \bar{\theta}_R^{\mathcal{A}} \gamma \cdot \mathring{T}^j \gamma_a \epsilon^k+\varepsilon_{jk} \xi_{\mathcal{A}}{}^i Y_a^j \bar{\psi}^{k \mathcal{A}} \epsilon_L \nonumber\\
    &+ \frac{1}{16} \xi_\mathcal{A}{}^i \bar{\psi}^{k \mathcal{A}} \gamma \cdot \mathring{T}_k \gamma_a \epsilon_R -\frac{1}{16} \xi_{\mathcal{A}}{}^i \bar{\psi}^{k \mathcal{A}} \gamma \cdot T^- \gamma_a \epsilon_k -\frac{1}{2} \xi_{\mathcal{A}}{}^i \bar{\theta}_R^{\mathcal{A}} \gamma_a \eta_L + \frac{1}{2} \varepsilon_{jk}\xi_{\mathcal{A}}{}^i \bar{\psi}^{k \mathcal{A}} \gamma_a \eta^j
    \\[5pt]
    \delta \xi_i{}^\mathcal{A}=&-\bar{\epsilon}_i\theta_R^{\mathcal{A}}+\varepsilon_{ij}\bar{\epsilon}^j\psi_L^{\mathcal{A}}-\varepsilon_{ij}\bar{\epsilon}_L\psi^{j \mathcal{A}} \\[5pt]
    \delta \psi_i{}^\mathcal{A}=&-\frac{1}{2}\gamma\cdot \mathcal{F}^{+\mathcal{A}}\epsilon_i-2\varepsilon_{ij}(-\cancel{D}\xi^{j\mathcal{A}}\epsilon_L+\xi_k{}^\mathcal{A} \slashed{Y}^k\epsilon^j)-\frac{1}{4}E_i\xi^{k\mathcal{A}} \epsilon_k+\frac{1}{2}\bar{\Lambda}_L\theta_L^\mathcal{A}\epsilon_i\nonumber\\
    &
    -\frac{1}{2}\varepsilon_{ik} \gamma_a\epsilon_L\bar{\Lambda}_R\gamma^a\Lambda_L\xi^{k\mathcal{A}} +\psi_R^{\mathcal{A}}u(\epsilon)_i+2\varepsilon_{ij}\xi^{j\mathcal{A}}\eta_R   \\[5pt]
    \delta \psi^\mathcal{A}_R=&-\frac{1}{2}\gamma\cdot \mathcal{F}^{+\mathcal{A}} \epsilon_{R}-2\varepsilon_{jk}\cancel{D}\xi^{k\mathcal{A}}\epsilon^j-\frac{1}{4}E\xi^{k\mathcal{A}}\epsilon_k+\frac{1}{2}\bar{\Lambda}_L\theta_L^\mathcal{A}\epsilon_R-\psi_j{}^{\mathcal{A}} u(\epsilon) ^j+2\varepsilon_{jk}\xi^{j\mathcal{A}}\eta^k  \\[5pt]
    \delta \theta_L^\mathcal{A}=&-2\cancel{D}\xi^{i\mathcal{A}} \epsilon_i-2\xi_j{}^\mathcal{A} \slashed{Y}^j\epsilon_R-\gamma^a\bar{\Lambda}_L\gamma_a\Lambda_R\xi^{j\mathcal{A}} \epsilon_j+\varepsilon_{ij}E^i\xi^{j\mathcal{A}} \epsilon_L+\frac{1}{4}\varepsilon_{ij}E\xi^{i\mathcal{A}} \epsilon^j-\bar{\Lambda}_R\psi_R{\mathcal{A}} \epsilon_L-2\xi^{i\mathcal{A}} \eta_i  \\[5pt]
 \delta \Tilde{C}_\mu{}^\mathcal{A}=&-i\bar{\epsilon}^i\gamma_\mu\psi_i{}^\mathcal{A}-i\bar{\epsilon}_L\gamma_\mu\psi_R^{\mathcal{A}} +2i(\bar{\epsilon}_L\psi_\mu^j-\bar{\epsilon}^j\psi_{\mu L} )\xi^{k\mathcal{A}}\varepsilon_{jk}+i\bar{\epsilon}_i\gamma_\mu\Lambda_L\xi^{i\mathcal{A}} +\text{h.c.} \\[5pt]
 \delta C_\mu{}^\mathcal{A}=&\bar{\epsilon}^i\gamma_\mu\psi_i{}^\mathcal{A}+\bar{\epsilon}_L\gamma_\mu\psi_R^{\mathcal{A}} +2(\bar{\epsilon}_L\psi_\mu^j-\bar{\epsilon}^j\psi_{\mu L} )\xi^{k\mathcal{A}}\varepsilon_{jk}-\bar{\epsilon}_i\gamma_\mu\Lambda_L\xi^{i\mathcal{A}} +\text{h.c.}\\[5pt]
 \delta B_{\mu \nu}{}^{\mathcal{A} \mathcal{B} }=& \frac{1}{2} C_{[\mu}{}^{(\mathcal{A}} \delta C_{\nu]}{}^{\mathcal{B})} + \frac{1}{2} \Tilde{C}_{[\mu} {}^{(\mathcal{A}}\delta \Tilde{C}_{\nu]}{}^{\mathcal{B})}  + ( \bar{\epsilon}_i \gamma_{\mu \nu} \theta_R^{(\mathcal{A}} \xi^{i \mathcal{B})} - \varepsilon_{ij} \bar{\epsilon}_L \gamma_{\mu \nu} \psi^{j (\mathcal{A}} \xi^{i \mathcal{B})} + \varepsilon_{ij} \bar{\epsilon}^j \gamma_{\mu \nu} \psi_L^{(\mathcal{A}} \xi^{i \mathcal{B})}\nonumber \\
    &-4 \xi_i^{(\mathcal{A}} \xi^{j \mathcal{B)}} \bar{\epsilon}^i \gamma_{[\mu} \psi_{\nu] j}+ 2 \xi^{i \mathcal{(A}} \xi_i{}^{\mathcal{B)}} \bar{\epsilon}_L \gamma_{[\mu} \psi_{\nu]R} +2\xi^{i \mathcal{(A}} \xi_i{}^{\mathcal{B)}} \bar{\epsilon}^j \gamma_{[\mu} \psi_{\nu] j} +\text{h.c} )\\[5pt]
  \delta E_{\mu \nu}=&  \frac{1}{2} \epsilon_{\mathcal{A} \mathcal{B}} \Tilde{C}^{\mathcal{B}}_{[\mu} \delta C^{\mathcal{A}}_{\nu]} + \frac{1}{2} \epsilon_{\mathcal{A} \mathcal{B}} C_{[\mu} {}^{\mathcal{A}}\delta \Tilde{C}_{\nu ]}{}^{\mathcal{B}} +\epsilon_{\mathcal{A} \mathcal{B}}  (  i \bar{\epsilon}_i \gamma_{\mu \nu} \theta_R^{\mathcal{B}} \xi^{i \mathcal{A}}  -i \varepsilon_{ij} \bar{\epsilon}_L \gamma_{\mu \nu} \psi^{j \mathcal{B}} \xi^{i \mathcal{A}} + i \varepsilon_{ij} \bar{\epsilon}^j \gamma _{\mu \nu} \psi_L^{\mathcal{B}} \xi^{i \mathcal{A}} \nonumber \\
    &-4i \xi^{i\mathcal{A}} \xi_{j}{}^{\mathcal{B}} \bar{\epsilon}_i \gamma_{[\mu} \psi_{\nu]}^j + 2i \xi^{i\mathcal{A}} \xi_i{}^{\mathcal{B}} \bar{\epsilon}^j \gamma_{[\mu} \psi_{\nu]j} + 2i \xi^{i \mathcal{A}} \xi_i{}^{\mathcal{B}} \bar{\epsilon}_L \gamma_{[\mu} \psi_{\nu] L} +\text{h.c})  
    \end{align}}
\end{subequations}
\end{widetext}

 \section{Conclusion and Future Directions} \label{conclusions}
 
 In this paper, we constructed a new dilaton Weyl multiplet for $\mathcal{N}=3$ conformal supergravity in four dimensions. This dilaton Weyl multiplet differs from the $\NN=3$ dilaton Weyl multiplet constructed in \cite{Adhikari:2024qxg} in its field contents. We constructed this new dilaton Weyl multiplet by coupling two vector multiplets with the standard Weyl multiplet and using their field equations to solve for some components of the auxiliary fields of the standard Weyl multiplet. This is in contrast with the dilaton Weyl multiplet constructed in \cite{Adhikari:2024qxg} where the coupling of one vector multiplet with the standard Weyl multiplet was used. As a result, in the new dilaton Weyl multiplet constructed in this paper, more components of the auxiliary fields of the standard Weyl multiplets become composite in contrast to the dilaton Weyl multiplet in \cite{Adhikari:2024qxg}. We refer to this new dilaton Weyl multiplet as a variant dilaton Weyl multiplet. One may wonder if one can use three vector multiplets coupled to the standard Weyl multiplet to construct yet another variant of $\NN=3$ dilaton Weyl multiplet. Since $\NN=3$ Poincar{\'e} supergravity is gauge equivalent to $\NN=3$ conformal supergravity that uses the standard Weyl multiplet and three compensating vector multiplets, this new variant might be gauge equivalent to a Poincar{\'e} supergravity multiplet without the necessity of any compensating multiplets. However, there is a subtlety in obtaining this new variant which is as follows. The number of components of fermionic field equations arising from the coupling of three vector multiplets to the standard Weyl multiplet would be $12\times 4=48$ whereas the number of fermionic auxiliary fields present in the standard Weyl multiplet is $9\times 4=36$. Hence, even after solving for all the auxiliary fields belonging to the standard Weyl multiplet, some more fermionic equations will still remain and it is not clear how one can solve them to obtain a completely off-shell dilaton Weyl multiplet. If one can resolve this subtlety, this can lead to a completely off-shell formulation of $\NN=3$ Poincar{\'e} supergravity.  
 

 In the dilaton Weyl multiplets constructed in this paper as well as in \cite{Adhikari:2024qxg}, coupling of on-shell $\NN=3$ vector multiplets to the standard Weyl multiplet has been used. However, recently the authors of \cite{Kuzenko:2025bud} have proposed a new matter multiplet, an on-shell non-linear multiplet, in $\NN=3$ conformal supergravity. This offers new approaches to the construction of dilaton Weyl multiplets in $\NN=3$ conformal supergravity.
 
 The Weyl multiplets for various amounts of supersymmetries are expected to be related to each other via supersymmetric truncations. In \cite{Adhikari:2024qxg}, it was argued that the dilaton Weyl multiplet constructed in their paper might be related to the $\NN=4$ dilaton Weyl multiplet constructed in \cite{Ciceri:2024xxf} via supersymmetric truncation. The $\NN=4$ dilaton Weyl multiplet constructed in \cite{Ciceri:2024xxf} used the coupling of one vector multiplet with the standard Weyl multiplet and hence it is plausible that it is related to the $\NN=3$ dilaton Weyl multiplet of \cite{Adhikari:2024qxg}. In a different work \cite{Adhikari:2024esl}, a variant $\NN=4$ dilaton Weyl multiplet was obtained by coupling another vector multiplet to the dilaton Weyl multiplet of \cite{Ciceri:2024xxf}. One may interpret this variant as being obtained from the coupling of two $\NN=4$ vector multiplets to the $\NN=4$ standard Weyl multiplet. Hence, one might wonder if this variant $\NN=4$ dilaton Weyl multiplet is related to the variant $\NN=3$ dilaton Weyl multiplet constructed in this paper via supersymmetric truncation. We reserve this investigation for future work.

 Often, supersymmetric truncation is helpful in identifying new multiplets in conformal supergravity with lesser amounts of supersymmetries from multiplets with higher amounts of supersymmetries. This approach has been exploited in \cite{Aikot:2024cne} in constructing new $\NN=2$ multiplets by using supersymmetric truncation of $\NN=3$ multiplets. In particular, it was observed that the $\NN=3$ standard Weyl multiplet truncates to an $\NN=2$ standard Weyl multiplet and an off-shell $\NN=2$ vector multiplet. Whereas an on-shell $\NN=3$ vector multiplet truncates to an on-shell $\NN=3$ vector multiplet and an on-shell scalar multiplet that can be identified with the $\mathcal{N} = 2$ hypermultiplet with a broken rigid $SU(2)$ and a non-trivial coupling to the off-shell vector multiplet. This non-trivial coupling of the scalar multiplet to the off-shell vector multiplet was further exploited to obtain a new $\NN=2$ multiplet which was referred to as the scalar-tensor multiplet. One may wonder if such a task of supersymmetric truncation on the dilaton Weyl multiplets constructed in this paper as well as that of \cite{Adhikari:2024qxg} offers any new insights into $\NN=2$ conformal supergravity. We leave this for future work.

As mentioned in the Introduction, conformal supergravity is often used as an intermediate step in the construction of matter-coupled Poincar{\'e} supergravity where the Weyl multiplets and the matter multiplets play an important role. The entire procedure goes by the name of superconformal multiplet calculus. The number of compensating matter multiplets required in this procedure varies with the kind of Weyl multiplets being used. For $\NN=3$ supergravity, three compensating multiplets are essential for the construction using a standard Weyl multiplet as seen in \cite{Hegde:2022wnb}. However, the number of compensating multiplets is expected to reduce when dilaton Weyl multiplets are used. Two compensating multiplets might become essential if the dilaton Weyl multiplet of \cite{Adhikari:2024qxg} is used whereas one would expect the use of one compensating multiplet if the variant dilaton Weyl multiplet of this paper is used. The Poincar{\'e} supergravity constructed using these different approaches might be related to each other via some kind of duality transformations which would be interesting to see. We leave these investigations to future work.

It has been seen in \cite{Ozkan:2013nwa} and \cite{Mishra:2020jlc} that upon Poincar{\'e} gauge fixing, a mapping exist between the dilaton Weyl multiplet and the Yang-Mills multiplet in the case of $\NN=1$ supergravity in five dimensions and $\NN=2$ supergravity in four dimensions. This mapping allowed for the supersymmetric completion of the Riemann squared term in the respective Poincar{\'e} supergravity theories. This combined with other curvature squared invariants obtained using the standard procedures allowed for the supersymmetrization of arbitrary curvature squared invariants in these cases. One may wonder if this procedure can be extended to the $\NN=3$ case using the dilaton Weyl multiplet constructed in this paper or that of \cite{Adhikari:2024qxg}. However, there might be a subtlety in this case. The $\NN=3$ vector multiplet is an on-shell multiplet and hence by extension, the $\NN=3$ Yang-Mills multiplet would also be on-shell whereas the $\NN=3$ dilaton Weyl multiplets are off-shell. Hence finding a mapping between the dilaton Weyl multiplet and the Yang-Mills multiplet for $\NN=3$ supergravity might be tricky. This will also happen in other cases such as $\NN=4$ supergravity in four dimensions or $\NN=2$ supergravity in five dimensions where the vector multiplets and hence by extension, the Yang-Mills multiplet are on-shell. Having said that, their exists off-shell formulations of the $\NN=3$ vector multiplet and the $\NN=3$ Yang-Mills multiplet in harmonic superspace \cite{Galperin:1984bu,Galperin:1985uw,Galperin:1986id}. Such off-shell formulations need an infinite number of auxiliary fields. One may try to find if their exists a mapping of the $\NN=3$ dilaton Weyl multiplet with such off-shell Yang-Mills multiplet upon Poincar{\'e} gauge fixing and whether such mapping can lead to a supersymmetrization of the Riemann squared invariant. In order to execute this task, we need to understand $\NN=3$ dilaton Weyl multiplet in harmonic superspace. We will try to address this in future.

\acknowledgments
We thank Chennai Mathematical Institute and IIT Ropar for their hospitality during the course of this work. The research of MM was supported by appointment to the Young Scientist Training (YST) Program at the APCTP through the Science and Technology Promotion Fund and Lottery Fund of the Korean Government (MM).
			
			\appendix

               \section{R-symmetry breaking} \label{decomposition}
In this appendix, the details of the decomposition of the fields of the $\mathcal{N}=3$ standard Weyl multiplet as well as the vector multiplets carrying $SU(3)$ representation into $SU(2)$ representations are given (See Table-\eqref{break0}).
\begin{widetext}
\begin{center}
\begin{table}[h!]
	\centering
	\begin{tabular}{ |p{4cm}|p{6cm}|}
		\hline
		$SU(3)$ irreps & $SU(2)$ irreps\\
			\hline
		\multicolumn{2}{|c|}{Fields of the standard Weyl Multiplet}\\
		\hline
$e_\mu^a(\bf{1})$&$e_\mu^a(\bf{1})$\\
  $\psi_\mu^I(\bf{3})$ & $\psi_\mu^i(\bf{2})$\;, $\psi_{\mu L} $$(\bf{1})$\\
  $V_\mu{}^I{}_J(\bf{8}) $ & $\mathring{v}_\mu{}^i{}_j(\bf{3})$\;, $Y_a^i(\bf{2})$\;, $\mathring{v}_\mu(\bf{1}).$\\ 
  $A_\mu (\bf{1}) $ & ${A}_\mu(\bf{1}) $\\
		$T_{ab}^{I}$  $(\bf{5})$& $\mathring{T}_{ab}^i$ $(\bf{2})$\;,\;$T_{ab}^+$  $(\bf{1})$\\
  $E_I(\bf{\bar{3}})$ & $E_i(\bf{{2}})$\;, $E$ (\bf{{1}})\\
  $D^I{}_J$ (\bf{8}) & ${\mathring{D}
  }^i{}_j$(\bf{3})\;, $\mathring{D}^i(\bf{2})$\;, $\mathring{D}(\bf{1})$.\\
 $ \zeta^I$ (\bf{3})& $\mathring{\zeta}^i(\bf{2})$\;, $\mathring{\zeta}_L(\bf{1})$\\
 $\Lambda_L$(\bf{1})& $\Lambda_L$(\bf{1})\\
		$\chi_{IJ}$  $(\bf{\bar{6}})$&$\mathring{\chi}_{ij}$$(\bf{{3}})$\;, $\mathring{\chi}_i$$(\bf{{2}})$\;, $\chi_L (\bf{1})$.\\
		\hline
		\multicolumn{2}{|c|}{Fields of the vector Multiplet}\\
		\hline
$C_{\mu}{}^{\AAA}$ $(\bf{1})$& $C_{\mu}{}^{\AAA} $ $(\bf{1})$\\
$\xi_I{}^{\AAA}  $$(\bf{\bar{3}})$& $\xi_i{}^{\AAA}$$(\bf{{2}})$\;, $\xi^{\AAA}(\bf{1})$\\
$\psi_I{}^{\AAA}(\bf{\bar{3}}) $& $\psi_i{}^{\AAA} (\bf{{2}})$\;, $\psi_R^{\AAA}(\bf{1}) $\\
$\theta_L^{\AAA}(\bf{1}) $ & $\theta_L^{\AAA}(\bf{1}) $\\
 \hline
	\end{tabular}
	\caption{Decomposition of $SU(3)$ to $SU(2)$}
	\label{break0}	
\end{table} 
\end{center}
\end{widetext}
The relations between the original fields appearing in the LHS of the table and the decomposed fields appearing in the RHS of the table are given in \eqref{1relations}.
\begin{align}\label{1relations}
&\psi_\mu^i = \psi_\mu^i\;, \quad \psi_\mu^3= \psi_{\mu L}\;, \nonumber\\
&V_{\mu}{}^{i}{}_j + \frac{1}{2} \delta^i_j V_{\mu}{}^{3}{}_{3} = \mathring{v}_{\mu}{}^{i}{}_{j}\;, \quad -iV_{\mu}{}^{3}{}_{3} = \mathring{v}_{\mu}\;,\nonumber\\
&V_{a}{}^i{}_{3} - \frac{1}{2} u(\psi_a)^i = Y_a^i\;, \nonumber\\
&T^i_{ab} = \mathring{T}^i_{ab}, \quad T^3_{ab} = T^{+}_{ab}\;, \nonumber\\
&E_i = E_i\;, \quad E_3 =E\;,\nonumber\\
&D^{i}{}_{j} - \frac{1}{2}\delta^i_j D^{3}{}_{3} = \mathring{{D}}^i{}_j\;, \quad D^3{}_3 = \mathring{D}\;, \quad D^{i}{}_3 = \mathring{D}^i\;,\nonumber \\
&\zeta_i = \mathring{\zeta}_i\;, \quad \zeta_3 = \mathring{\zeta}_R\;, \nonumber \\
& \chi_{ij}= \mathring{\chi}_{ij}\;, \quad \chi_{i3}= \mathring{\chi}_i\;, \quad \chi_{33} = \chi_L\;, \nonumber \\
&\xi_i{}^{\AAA} = \xi_i{}^{\AAA}\;, \quad \xi_3^{\AAA} = \xi^{\AAA}\;, \nonumber \\
& \psi_i{}^{\AAA} = \psi_i{}^{\AAA}\;, \quad \psi_3^{\AAA} = \psi_R^{\AAA}\;.
\end{align}

The reality properties of some of the decomposed fields is inherited from the reality properties
of the original fields and are given as below:
\begin{align}
(\mathring{v}_\mu{}^i{}_j)^*&=-\mathring{v}_\mu{}^j{}_i\nonumber\\
(\mathring{v}_\mu)^*&=\mathring{v}_\mu\nonumber\\
(\mathring{D}^i{}_j)^*&=\mathring{D}^j{}_i \nonumber\\
(\mathring{D})^*&=\mathring{D}
\end{align}
\bibliography{references}
\bibliographystyle{utphys}
\end{document}